\documentclass[letterpapper,12pt]{article}

\usepackage[
  top=2.6cm,
  bottom=2.6cm,
  left=2.0cm,
  right=2.0cm
]{geometry}

\usepackage{graphicx}
\usepackage{amsmath}
\usepackage{authblk}
\usepackage[authoryear,longnamesfirst]{natbib}
\usepackage{indentfirst,csquotes}
\usepackage{booktabs}
\usepackage{setspace}
\usepackage{lipsum}
\usepackage[colorlinks=true,
            linkcolor=blue,
            citecolor=blue,
            urlcolor=blue]{hyperref}

\begin{document}
\title{\Large\textbf{Development and Characterization of Low-Scattering Vanadium Nanoparticle Targets\\ for Short-Range Interaction Searches}} 
\date{}
\author[1]{\large Masayuki Hiromoto}
\author[1]{\large Tatsushi Shima}
\author[2]{\large Christopher C. Haddock}
\author[3]{\large Katsuya Hirota}
\author[4,5]{\large Masaaki Kitaguchi}
\author[1]{\large Ryota Kondo}
\author[6]{\large Rintaro Nakabe}
\author[1]{\large Kenji Mishima}
\author[4]{\large Hirohiko M. Shimizu}
\author[1]{\large Yuki Yoshikawa}            
\author[7]{\large Tamaki Yoshioka}
\affil[1]{\large\itshape Research Center for Nuclear Physics (RCNP), Osaka University, 10-1 Mihogaoka, Ibaraki, Osaka 567-0047, Japan}
\affil[2]{\large\itshape Phase III Physics, Los Angeles, 90015, CA, USA}
\affil[3]{\large\itshape Institute of Materials Structure Science, High Energy Accelerator Research Organization (KEK), Tokai, 319-1106, Ibaraki, Japan}
\affil[4]{\large\itshape Department of Physics, Nagoya University, Nagoya, 464-8602, Aichi, Japan}
\affil[5]{\large\itshape Kobayashi-Maskawa Institute for the Origin of Particles and the Universe (KMI), Nagoya University, Nagoya, 464-8602, Aichi, Japan}
\affil[6]{\large\itshape Nuclear Science Research Institute, Japan Atomic Energy Agency (JAEA), Tokai, 319-1195, Ibaraki, Japan}
\affil[7]{\large\itshape Faculty of Arts and Science, Kyushu University, Fukuoka, 819-0395, Fukuoka, Japan}
\maketitle

\let\thefootnote\relax

\begin{abstract}
We developed high-purity vanadium-based nanoparticle targets for neutron scattering experiments aimed at exploring gravity-like short-range new interactions in the submicron regime. Vanadium and V-Ni nanoparticles were fabricated using top-down and bottom-up methods and quantitatively characterized by SEM-EDS, ICP-AES, NDIR and SAXS. Through the performance tests, an RF thermal plasma method was found to be the best from viewpoints of the reproducibility, dispersion of the radius, and contamination of metallic elements. The oxygen incorporation during fabrication was quantified, and its impact on the effective coherent scattering length was evaluated, leading to a minimum average coherent scattering length of $\mathrm{0.719(23)\,fm}$, comparable to that of natural vanadium. These results demonstrate that vanadium-based nanoparticle targets with controlled composition and nanostructure can be systematically designed and fabricated to suppress nuclear scattering backgrounds, thereby enabling experimentally viable coherent neutron scattering measurements for short-range interaction searches.
\end{abstract} 

\bigskip
\section{Introduction}
Elastic scattering experiments using low-energy neutrons are powerful experimental probes for studying gravity-like interactions~\cite{ARKANIHAMED1998263, PhysRevD.108.055005, Fujii:1971vv, Sponar:2021aa} in the submicron range.
Over the past decade, small-angle neutron scattering (SANS)~\cite{PhysRevD.97.062002,PhysRevLett.114.161101} and interferometry experiments~\cite{osti_1850182} have placed increasingly stringent constraints on Yukawa-type deviations from the inverse-square law at nanometer length scales.
A particularly promising strategy to further enhance sensitivity is to exploit coherent neutron scattering from nanostructured targets~\cite{PhysRevD.108.055005, Hiromoto:2021aa}. When the target size is comparable to the interaction range, the scattering amplitudes from individual nuclei add coherently, potentially amplifying the signal from hypothetical short-range forces by many orders of magnitude. This coherent enhancement makes nanoparticle-based targets attractive for next-generation searches.
However, the same coherent mechanism also amplifies ordinary nuclear scattering, which constitutes the dominant background in neutron experiments. The coherent nuclear scattering intensity scales with the square of the effective scattering length and with the number of nuclei contained in each nanoparticle. Therefore, minimizing the average coherent nuclear scattering length of the target material is essential for realizing the full advantage of coherent enhancement.
In this context, vanadium-based materials are particularly attractive because natural vanadium possesses one of the smallest coherent nuclear scattering lengths among all elements. Furthermore, null-matrix alloy concepts allow additional suppression of the effective scattering length by carefully tuning alloy composition. Nevertheless, the practical realization of such nanoparticle targets requires strict control of chemical purity, oxidation, and particle-size uniformity, as even small contaminations can significantly modify the effective scattering length.
In the present work, we establish design criteria for low-background nanoparticle targets suitable for coherent neutron scattering searches of short-range interactions. We fabricate vanadium and V-Ni nanoparticles using both top-down and bottom-up methods, and we quantitatively evaluate their structural and chemical properties. Based on comprehensive characterization and neutron scattering calculations, we demonstrate that RF thermal plasma synthesis enables reproducible production of nanoparticle targets with controlled composition and predictable nuclear scattering properties, satisfying the key experimental requirements for precision searches of gravity-like interactions.  The nuclear scattering cross section was calculated from the elemental composition ratios contained in the nanoparticles, and the smallest coherent nuclear scattering length of the fabricated nanoparticles was calculated to be $\mathrm{0.719(23)\,fm}$.

\section{Target Material Design}\label{Sec.2}
\subsection{Requirements for neutron coherent scattering}
\label{Sec.2.1}
The advantage of using nanoparticles as the target for new interactions searches is that the coherent scattering cross section can be significantly enhanced thanks to the coherent effect. The differential cross section for coherent neutron scattering from an isolated particle, including the scattering lengths $b_\mathrm{Y}(q)$ of a hypothetical short-range new interaction~\cite{Frank_2004}, can be expressed within the first Born approximation as
\begin{equation}
\label{eq1}
\left(\frac{d\Sigma}{d\Omega}\right)_{\mathrm{coh}}
= N(R)\,F^2(q,R)\,
\left(b_{\mathrm{coh}} + b_\mathrm{E}(q) + b_\mathrm{M}(q) + b_\mathrm{Y}(q)\right)^2 ,
\end{equation}
Here, $N(R)$ is the number density of nanoparticles of radius $R$, $b_{\mathrm{coh}}$ is the average coherent nuclear scattering length of atoms contained in the nanoparticles that contribute to the coherent nuclear scattering. 
The electromagnetic scattering length $b_\mathrm{E}(q)$ is the scattering due to the imbalance of the internal charge distribution of the neutron. Since the neutron is charge-neutral to a high degree, the permanent electric dipole moment is zero or negligible~\cite{10.1093/ptep/ptac097}, but since it is a composite particle, it can have a finite dielectric polarization. Neutrons are scattered by their internal charge distribution due to their interaction with the electron electric field of an atom or the electric field of an atomic nucleus. The former is calculated as $Zb_\mathrm{ne}(1-f_\mathrm{ne}(q))$ using the electron form factor $f_\mathrm{ne}(q)$~\cite{PhysRevC.56.2229}, and the scattering length is $b_\mathrm{ne}\approx 10^{-3}\,\mathrm{fm}$. In the $q\approx 0.03\sim 0.1\,\mathrm{nm^{-1}}$ region used for measuring nano-targets, the value of the electron form factor becomes very small in proportion to $q^2$, and the scattering length due to this interaction is approximately $10^{-7}\sim 10^{-6}\,\mathrm{fm}$. The latter is a Coulomb interaction with the nuclear electric field~\cite{SEARS1986281}, and the main scattering length for describing it does not include a $q$-dependence itself, which is estimated to be about $10^{-4}\,\mathrm{fm}$ based on the nuclear radius and the neutron dielectric polarizability $\alpha_n\approx 10^{-3}\,\mathrm{fm^3}$~\cite{10.1093/ptep/ptac097}. The principal scattering length describing this is constant and does not include $q$ dependence, so it can be considered as the principal electromagnetic scattering length $b_\mathrm{E}(q)\approx b_\mathrm{E}$.\\
The magnetic scattering length $b_\mathrm{M}(q)$ arises from the interaction between the neutron's magnetic moment and the atomic electric field or the magnetic moment of the target material, both of which are neutron spin-dependent terms. The former, called the Schwinger term~\cite{PhysRev.73.407}, depends on the neutron spin and is characterized on the atomic length scale. The scattering intensity attenuates approximately proportionally to $q$, and in the small-angle region ($q\ll0.1\,\mathrm{nm^{-1}}$) used to measure nano-targets, is less than $10^{-4}\,\mathrm{fm}$. The latter reflects the spatial distribution of magnetic moments within nanoparticles as dipolar interactions and is modulated by the particle shape factor. In the current SANS measurement range, the magnetic form factor is expected to approach unity for small $q$ and decrease only when $q$ exceeds the reciprocal of the particle size. The scattering length due to this effect, explained in Sec.~\ref{Sec.2.2} below based on the characteristics of the target elements, is estimated to be on the order of $10^{-4}\,\mathrm{fm}$ or less.
$F(q,R)=\int\rho(r)\exp(iQ\cdot r) d^3r$ is the form factor of the nanoparticle of radius $R$. $F(q,R)$ is defined by the Fourier transform of the scatterer density, and its $q$ dependence is determined by the shape of the scatterer. To isolate the contribution of the hypothetical interaction, we consider the ratio of the scattering lengths for the new interaction and the nuclear elastic scattering. From the equations containing the respective scattering lengths discussed above, when using unpolarized neutrons, the ratio of the experimental and the theoretical coherent scattering cross sections, the former may 
contain the effect of the new interaction and the latter is given by the nuclear scattering alone, is given by the following equation.
\begin{equation}
\begin{aligned}
\label{eq2}
R(q)&=\frac{\left(d\Sigma/d\Omega\right)_{\mathrm{data}}}
{\left(d\Sigma/d\Omega\right)_{\mathrm{cal}}},\\
&=\frac{b_{\mathrm{coh}}^2 + 2b_{\mathrm{coh}}b_\mathrm{E} + 2b_{\mathrm{coh}}b_\mathrm{Y}(q) + b_\mathrm{M}(q)^2 + b_\mathrm{Y}(q)^2}{b_{\mathrm{coh}}^2},\\
&\approx\left(1 + \frac{b_Y(q)}{b_{\mathrm{coh}}}\right)^2\approx1+\frac{2b_Y(q)}{b_{\mathrm{coh}}} .
\end{aligned}
\end{equation}

\begin{figure}
  \centering
  \includegraphics[width=0.6\textwidth]{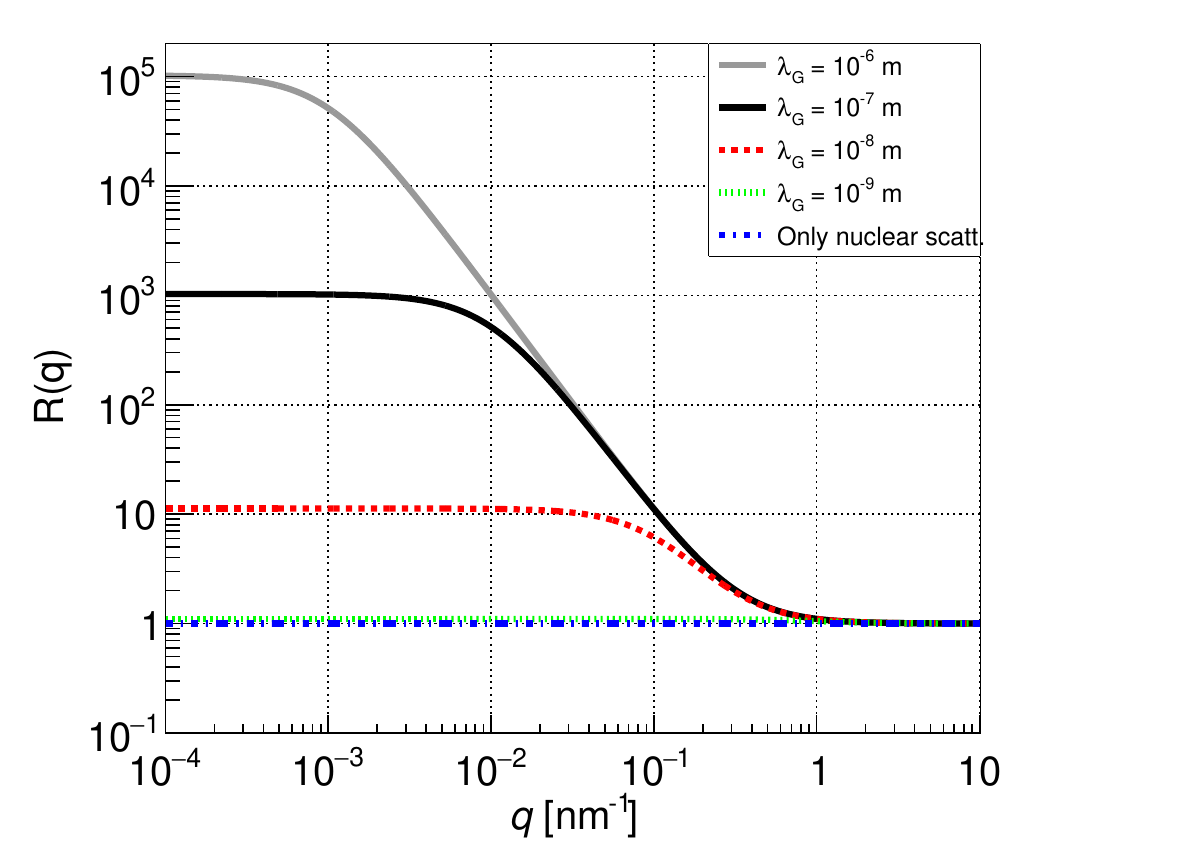}
  \caption{The $q$ dependence of the new interaction for various values of the range $\lambda_\mathrm{G}$, which appears in the scattering intensity ratio $R(q)$ to nuclear scattering. The calculation is based on the coherent nuclear scattering length of vanadium atoms [$b_\mathrm{coh}=-0.555\,\mathrm{nm}$] and the new interaction with a relative coupling constant of $\alpha=10^{22}$.}
  \label{fig:Rq_est}
\end{figure}

Here, $\left(d\Sigma/d\Omega\right)_{\mathrm{data}}$ and $\left(d\Sigma/d\Omega\right)_{\mathrm{cal}}$ are the $q$ distributions calculated from the measured small-angle neutron scattering data and the nuclear scattering cross section, respectively. In the following discussion, we assume that the scattering length due to electromagnetic interactions is negligibly smaller than the coherent nuclear scattering length. Furthermore, we assume that the new interaction is much weaker than the nuclear interaction, so we assume $|b_{\mathrm{coh}}| \gg |b_{\mathrm{Y}}(q)|$. In this case, the $q$ dependence of the new interaction is shown in Fig.~\ref{fig:Rq_est} for various values of the range $\lambda_\mathrm{G}$.

The cross section for each interaction is maximized in the region corresponding to the inverse of the momentum transfer. By utilizing scattered waves from atomic ensembles contained in a target of a size corresponding to each interaction length, coherent scattering can increase the cross section of the target, leading to an improvement of the statistical uncertainty of the measurement. On the other hand, the use of coherent scattering also increases nuclear scattering, which is the main scattering factor for neutrons, and this increases the background noise. 
When searching for new interactions using coherent scattering, the sensitivity can be improved by employing targets with a reduced average coherent nuclear scattering length.

\subsection{Vanadium and V-Ni alloy}
\label{Sec.2.2}
\begin{figure}
\centering
    \centering
    \includegraphics[width=0.6\textwidth]{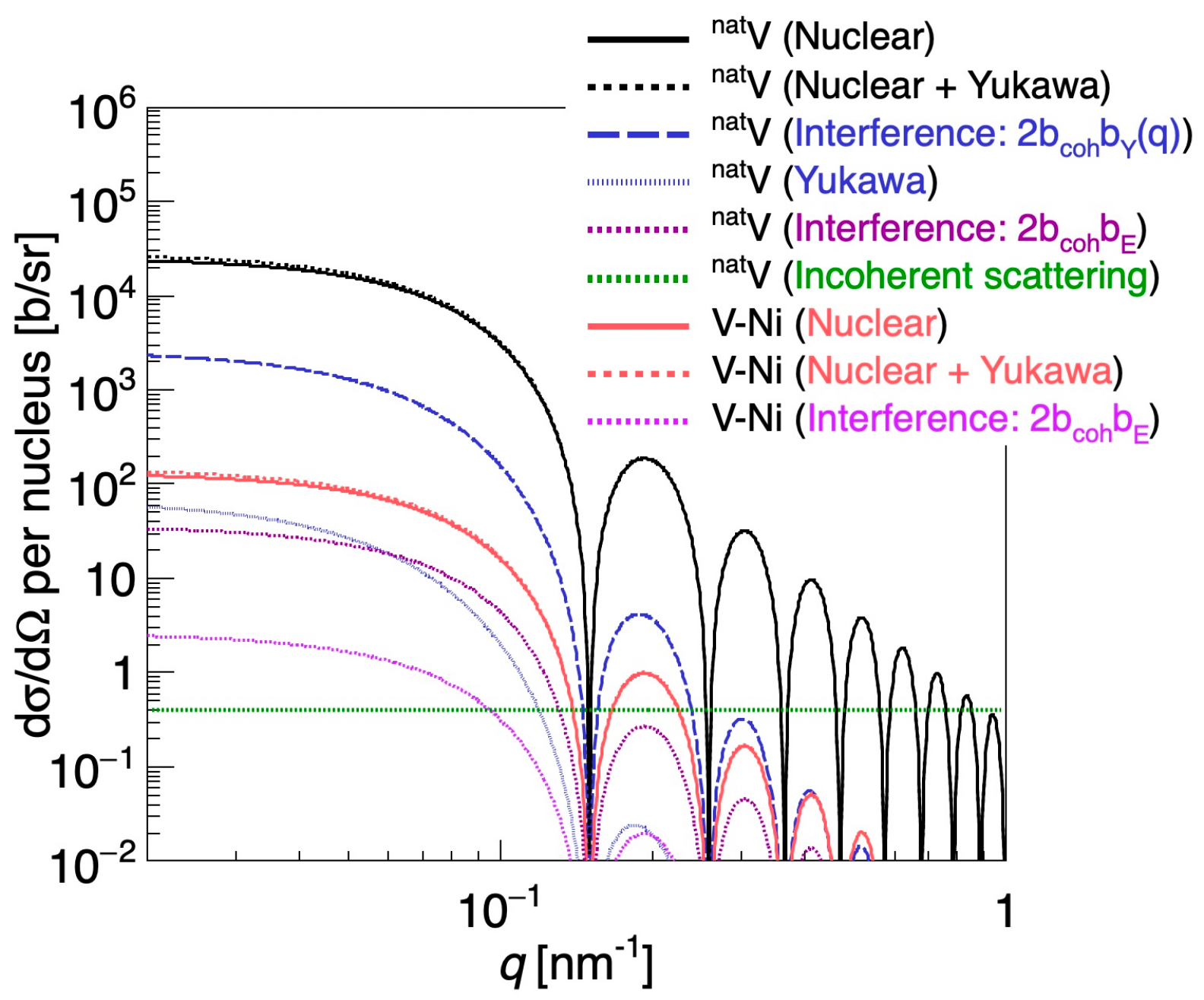}
    \caption{Comparison of neutron coherent scattering differential cross sections including new interactions with the targets made of natural vanadium and V-Ni alloys. Single hard sphere of $R=\mathrm{30\,nm}$ was assumed as the shape of the target nanoparticle. The interaction length of the new interaction were set to $\lambda_\mathrm{G}=10\,\mathrm{nm}$. Under the condition $|b_{\mathrm{coh}}|\gg|b_{\mathrm{Y}}(q)|$, the value of $\alpha$, which is the deviation of nuclear coherent scattering by $10\,\%$, was set, taking into account the influence of new interactions. The coupling constant of new interaction is set to $\alpha=1\times10^{20}$ for $\mathrm{^{nat}V}$ and $\alpha=7\times10^{18}$ for V-Ni.}
    \label{fig:Vnull_1}
\end{figure}

To suppress coherent nuclear scattering, we selected natural vanadium and V-Ni alloys as materials for fabricating nanoparticles. The coherent nuclear scattering length of natural vanadium has been measured as $b_\mathrm{V}=\mathrm{-0.555(3)\,fm}$ using a state-of-the-art neutron interferometer~\cite{PhysRevLett.132.023402}, which is the smallest among all the natural elements. Bulk metallic vanadium exhibits paramagnetic properties, with spin and orbital susceptibilities of less than $10^{^{-4}}\,\mathrm{emu/mol}$ at room temperature~\cite{Yasui1971CalculationsOO}, respectively. In the presence of an external magnetic field of $1\,\mathrm{mT}$, the magnetic scattering amplitude is expected to be $b_\mathrm{M}(q)\approx10^{-7}\,\mathrm{fm}$. In this case, the largest contributions from the electromagnetic interactions are estimated to be the scattering length $b_\mathrm{E}$ due to induced polarization and the interference term due to the coherent nuclear scattering length. Furthermore, by fabricating nanoparticles using null matrix alloys~\cite{JH_Smith_1968, met5042340, doi:10.1021/acsaenm.4c00553}, we expect improvement in the sensitivity of exploring new interactions compared to natural vanadium. V-Ni alloy is an alloy in which nickel, which has a coherent nuclear scattering length of $\mathrm{+10.3\,fm}$, is added to vanadium.
The average coherent nuclear scattering length $\langle b_\mathrm{coh} \rangle$ of the alloy made of two elements A and B is given by $\langle b_\mathrm{coh} \rangle=kb_\mathrm{A}+(1-k)b_\mathrm{B}$, where $b_\mathrm{A}$ and $b_\mathrm{B}$ are the scattering lengths of the elements A and B, respectively, and $k$ is the composition ratio of elements with positive scattering lengths. In the case of V-Ni alloy, Ni with $\mathrm{5.8\,wt\%}$ concentration provide $\langle b_\mathrm{coh} \rangle$ close to zero. Figure~\ref{fig:Vnull_1} shows a comparison of the differential cross sections for coherent neutron scattering, including new interactions, between natural vanadium and V-Ni alloy. The composition ratio of V-Ni alloys was set to $\pm\mathrm{0.5\,wt\%}$, assuming realistic uncertainty in the nickel content, and calculations were performed using $\langle b_\mathrm{coh}\rangle = 0.00\pm0.04\,\mathrm{fm}$. Figure~\ref{fig:Vnull_1} shows calculations for a single hard sphere with a radius of $\mathrm{30\,nm}$, with the conditions for the new interaction set at $\lambda_\mathrm{G}=10\,\mathrm{nm}$, $\alpha=1\times10^{20}$ for natural vanadium and $\alpha=7\times10^{18}$ for V-Ni alloy. 

\begin{figure}
    \centering
    \includegraphics[width=0.5\textwidth]{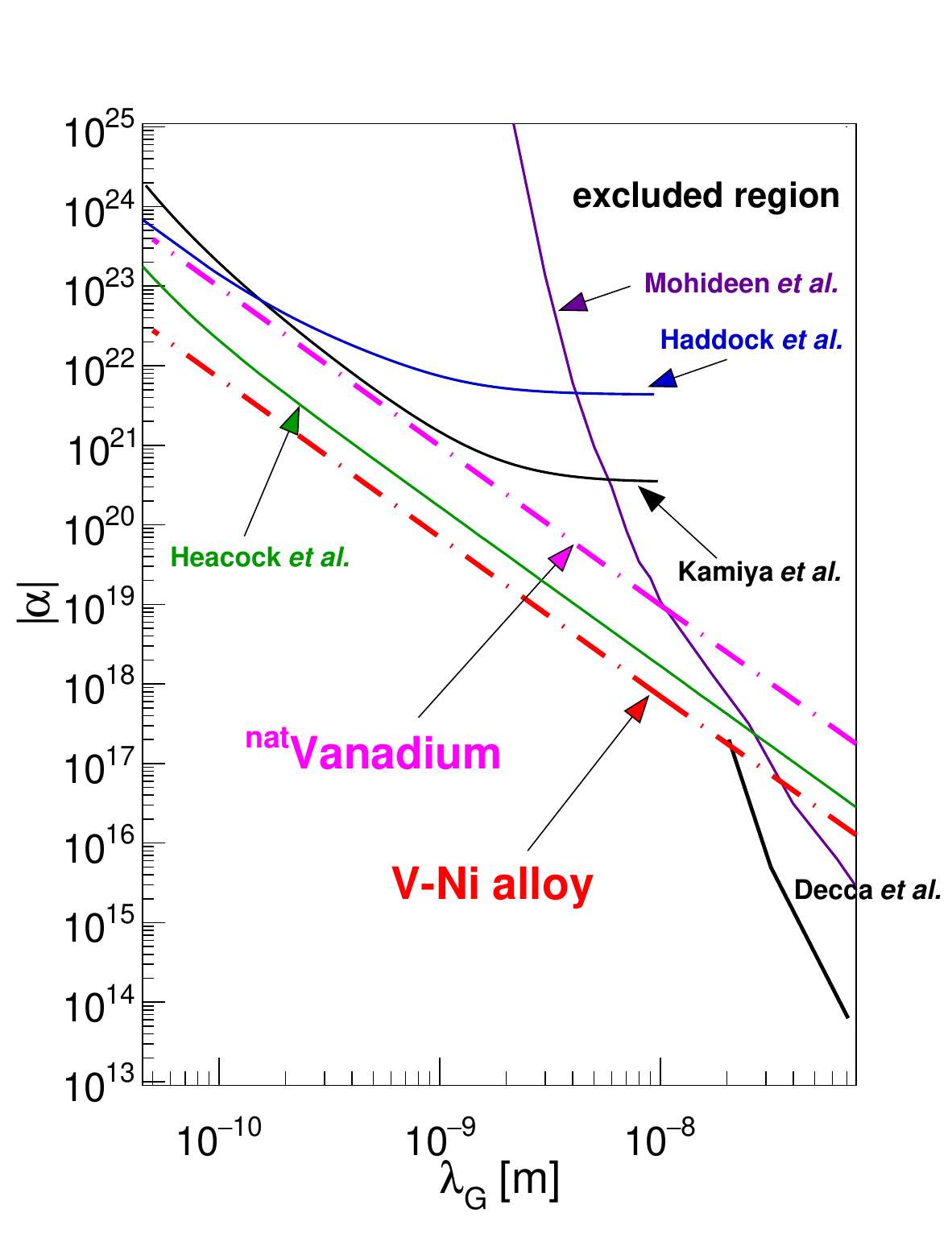}
    \caption{Predicted exclusion region at $\mathrm{95\,\%}$ C.L. of new interactions that can be explored by V and V-Ni nanoparticles.The estimation results assumed that a $1\,\mathrm{\%}$ deviation in the $R(q)$ ratio could be verified with $95\,\mathrm{\%}$ confidence.}
    \label{fig:Vnull_2}
\end{figure}

From the ratio given in Eq.~(\ref{eq2}), we set the value of $\alpha$, which is the deviation of $10\,\%$ nuclear coherent scattering under the condition $|b_{\mathrm{coh}}|\gg|b_{\mathrm{Y}}(q)|$, taking into account the influence of the new interaction. The use of V-Ni nanoparticles reduces coherent nuclear scattering by two orders of magnitude compared to natural vanadium, improving the search sensitivity for $\alpha$, which is more than an order of magnitude smaller than the new interaction cross section. Figure~\ref{fig:Vnull_2} shows the search region for new interactions that can be achieved when verifying $\alpha$ where the cross section ratio of the new interaction to the nuclear scattering cross section is $1\,\%$ with a confidence level of $95\,\%$. 
\clearpage

The reduction in nuclear scattering improves the range of $\alpha$ that can be investigated. Compared to the $\lambda_\mathrm{G}=10\,\mathrm{nm}$ region excluded in previous experiments on rare gas atomic targets, the expected improvement is approximately $\mathcal{O}(10^1)$ times for natural vanadium and approximately $\mathcal{O}(10^3)$ times for V-Ni alloys. Note that while the interference term with induced polarization contributes approximately $\mathrm{0.1\,\%}$ compared to nuclear scattering in natural vanadium, it may fluctuate by several $\mathrm{\%}$-fold in V-Ni alloys. However, since the primary effect of induced polarization is independent of $q$, this is unlikely to affect the evaluation of the $q$-dependence of the hypothetical new interaction. However, the fabrication of these nanoparticle targets requires high purity and structural uniformity to suppress changes in the coherent nuclear scattering length due to contamination with other elements. In particular, vanadium, which is generally prone to oxidation, must be fabricated in an inert gas atmosphere. 
The coherent nuclear scattering length of oxygen is as large as $b_{\mathrm{coh}}=+5.805\,\mathrm{fm}$. $\langle b_\mathrm{coh} \rangle \approx 0$ occurs when the oxygen concentration in vanadium is $\mathrm{3\,wt\%}$. Furthermore, when the oxygen concentration is $\mathrm{6\,wt\%}$ or higher, $\langle b_\mathrm{coh} \rangle\geq b_\mathrm{V}$, exceeding that of natural vanadium.
Therefore, to obtain the coherent nuclear scattering cross section smaller than the one with pure vanadium, oxygen concentration must be kept below $\mathrm{6\,wt\%}$.
This requirement imposes practical constraints on both the manufacturing method and the handling environment. To fabricate metal nanoparticles with arbitrary composition ratios, either the top-down method, in which raw materials with adjusted composition ratios are converted into nanoparticles, or the bottom-up method, in which nanostructures are constructed from atomic levels, are considered. We fabricated natural vanadium and V-Ni alloy nanoparticles using those two methods.

\section{Fabrication and characterization}\label{}
\subsection{Fabrication procedure}
In this section, we describe the test methods for the preparation of natural vanadium and V-Ni nanoparticles using two different methods, namely, top-down and bottom-up methods.

\subsubsection{Jet mill method}
\begin{figure}
    \centering
    \includegraphics[width=0.6\textwidth]{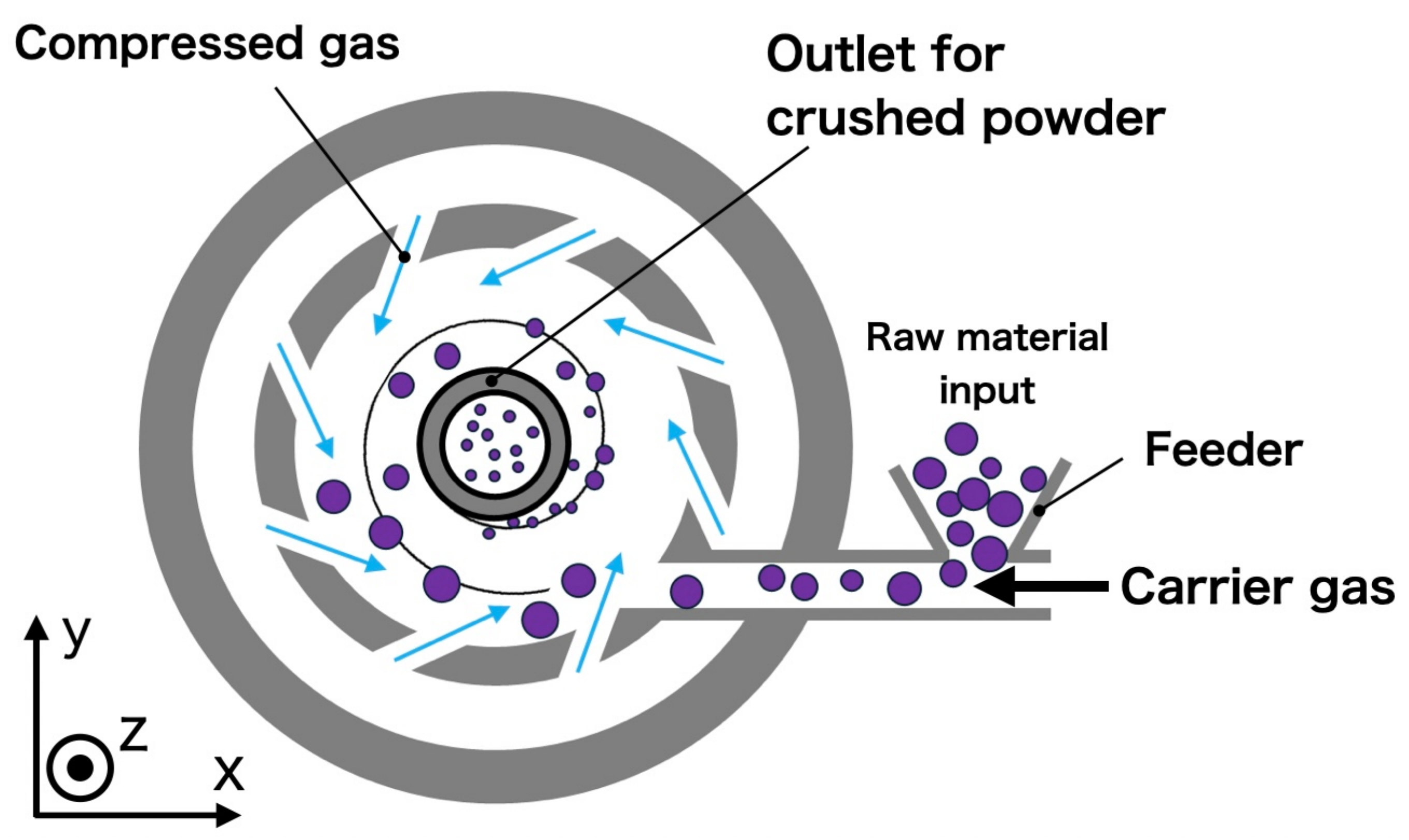}
\caption{Jet mill set-up of prototype test.}
\label{fig:JetMill}
\end{figure}

Bulk V-Ni alloy material was prepared and subjected to a nanoparticle production test using the top-down milling method. When using bulk materials as the raw material, contamination due to oxidation of the raw material is limited to the surface because particles are covered with thin passive layer of vanadium oxide which avoid further oxidation of inner part. However, contamination from the equipment is unavoidable with the milling method. Therefore, we used the jet milling method, in which the contamination effect is known to be rather low compared to the other
methods. The milling machine used was the jet mill (NJ-50) of Aisin Nano Technologies Co., Ltd.
As shown in Fig.~\ref{fig:JetMill}, the entire device consists of a powder feeder, a chamber made of zirconia for jet mill, and a sample collection filter. 
The raw powder is transported by helium carrier gas from the powder feeder into the zirconia jet mill.  In a flow of nitrogen gas, large grains of the raw material collide with each other to produce smaller nanoparticles, which are collected by a sample collection filter via a hose from the outlet in the center of the device.
The V-Ni alloy was produced in an arc melting furnace using an alloy containing $\mathrm{4.9\,wt\%}$ of natural nickel, in the form of a square plate with a thickness of $\mathrm{250\,\mu m}$ and an area of $\mathrm{100\,mm\times100\,mm}$. The composition ratio of the V-Ni alloy during production is shown in Table~\ref{Table:VNifoil_ana}, and the calculated value from the composition ratio was $\langle b_\mathrm{coh}\rangle=-0.09\,\mathrm{fm}$. 
\begin{table}
    \centering
    \normalfont
    \caption{Elemental compositions in V-Ni alloy foil in unit of $\mathrm{wt\%}$.}
    \label{Table:VNifoil_ana}
    \setlength{\tabcolsep}{0.05cm}
    \renewcommand{\arraystretch}{1.0}
\begin{tabular}{c c c c c c c c c c c}
\toprule
\multicolumn{1}{c}{} & \multicolumn{1}{c}{V} & \multicolumn{1}{c}{Ni} & \multicolumn{1}{c}{Al} & \multicolumn{1}{c}{Si} & \multicolumn{1}{c}{Fe} & \multicolumn{1}{c}{Mo} & \multicolumn{1}{c}{C} & \multicolumn{1}{c}{O} & \multicolumn{1}{c}{N}\\\midrule
\begin{tabular}[c]{@{}l@{}} Foil\# \end{tabular}
& \begin{tabular}[c]{@{}c@{}} Bal. \end{tabular}
& \begin{tabular}[c]{@{}c@{}} 4.81 \end{tabular}
& \begin{tabular}[c]{@{}c@{}} 0.003 \end{tabular}
& \begin{tabular}[c]{@{}c@{}} 0.015 \end{tabular}
& \begin{tabular}[c]{@{}c@{}} 0.003 \end{tabular}
& \begin{tabular}[c]{@{}c@{}} 0.001 \end{tabular}
& \begin{tabular}[c]{@{}c@{}} 0.008 \end{tabular}
& \begin{tabular}[c]{@{}c@{}} 0.016 \end{tabular}
& \begin{tabular}[c]{@{}c@{}} 0.013 \end{tabular}\\\bottomrule
  \end{tabular}
\end{table}

To operate the jet-mill process efficiently, it is necessary to make the size of the raw material smaller than $500\,\mathrm{\mu m}$. Therefore, a vanadium plate was cut into grains with scissors and sieved out to obtain grains with the size less than $500\,\mathrm{\mu m}$.
The jet mill grinding process was carried out in a circulating glove box, with the oxygen concentration and dew point recorded. The test environment was maintained at an oxygen concentration of $\mathrm{0.0\pm0.1\,wt\%}$ and a dew point of approximately $\mathrm{-30\,^\circ C}$.  At the carrier gas pressure of $\mathrm{2.0\,MPa}$ and the internal pressure of the jet mill of between $\mathrm{0.5}$ and $\mathrm{1.5\,MPa}$.
the throughput of the raw material was $\mathrm{60\,g/h}$.
The result of the production test is summarized in Table~\ref{Table:Jetmill_condi}, where $m_\mathrm{in}$ is the amount of powder fed into the jet mill, $P_\mathrm{in}$ is the pressure inside the jet mill, and $m_\mathrm{box}$ and $m_\mathrm{inside}$ are the sample amounts recovered from the collection box and jet mill.

\begin{figure}
    \centering
    \includegraphics[width=0.46\textwidth]{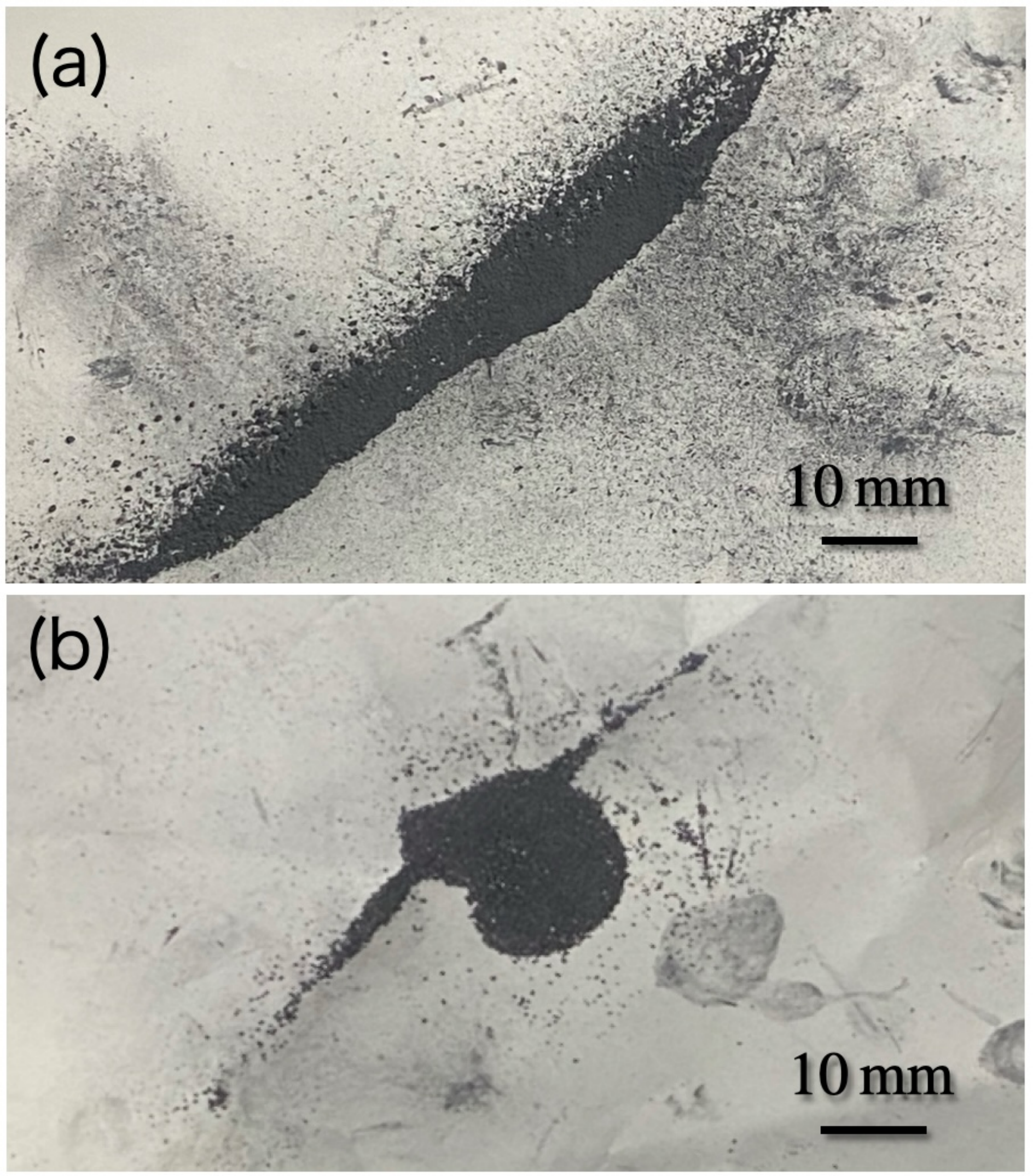}
\caption{V-Ni alloy powder after crushing. (a) Sample powder remaining in the jet mill, (b) Sample powder in the collection box.}
\label{fig:JetMill_after}
\end{figure}

\begin{table}
    \centering
    \normalfont
    \caption{Test conditions for V-Ni alloy crushing using jet mill.}
    \label{Table:Jetmill_condi}
    \setlength{\tabcolsep}{0.15cm}
    \renewcommand{\arraystretch}{1.0}
\begin{tabular}{lclccc}
\toprule
Step & \multicolumn{1}{c}{$m_\mathrm{in}\,\mathrm{(g)}$} & \multicolumn{1}{c}{$P_\mathrm{in}\,\mathrm{(MPa)}$} &  \multicolumn{1}{c}{Time$\,\mathrm{(h)}$} & \multicolumn{1}{c}{$m_\mathrm{box}\,\mathrm{(g)}$} & \multicolumn{1}{c}{$m_\mathrm{inside}\,\mathrm{(g)}$} \\
\midrule
\#1 & 6.06 & 0.6 $\rightarrow$ 1.0   & 0.5  & 1.02  & $-$ \\
\#2 & 1.02 & 1.40 $\rightarrow$ 1.35 & 0.5  & 0.83  & 4.13 \\
\#3 & 4.80 & 1.50                    & 1.25 & 1.38  & 1.35 \\
\#4 & 2.73 & 1.1 $\rightarrow$ 0.4   & 1.0  & 1.06  & 1.09 \\
\#5 & 2.15 & 0.7 $\rightarrow$ 0.4   & 0.5  & 1.06  & 1.05 \\
\bottomrule
\end{tabular}
\end{table}
During the crushing process, some crushed material remained inside the machine and some was collected in a collection box. After collection, these were re-introduced into the jet mill and crushed again for five times. As shown in Fig.~\ref{fig:JetMill_after}, the appearance of the samples after milling showed that the powder recovered from inside the jet mill appeared finer than that in the collection box. 
After the test, $\mathrm{1.01\,g}$ of powder was recovered from inside the jet mill. To classify the sample powder obtained from the jet mill, the sample powder was placed in an ethanol solution, dispersed using an ultrasonic vibrator, and filtered using a syringe equipped with a syringe filter with a mesh size of $\mathrm{0.45\,\mu m}$. The resulting classified solution was placed in an eggplant flask and dried using a rotary evaporator. The sample powder adhering to the eggplant flask was then collected with a metal spatula in a vacuum glove box, and $\mathrm{128\,mg}$ of classified $\mathrm{VNi_{jm}}$ sample powder was recovered.

\subsubsection{RF thermal plasma method}
\begin{figure}
    \centering
    \normalfont
    \includegraphics[width=0.6\textwidth]{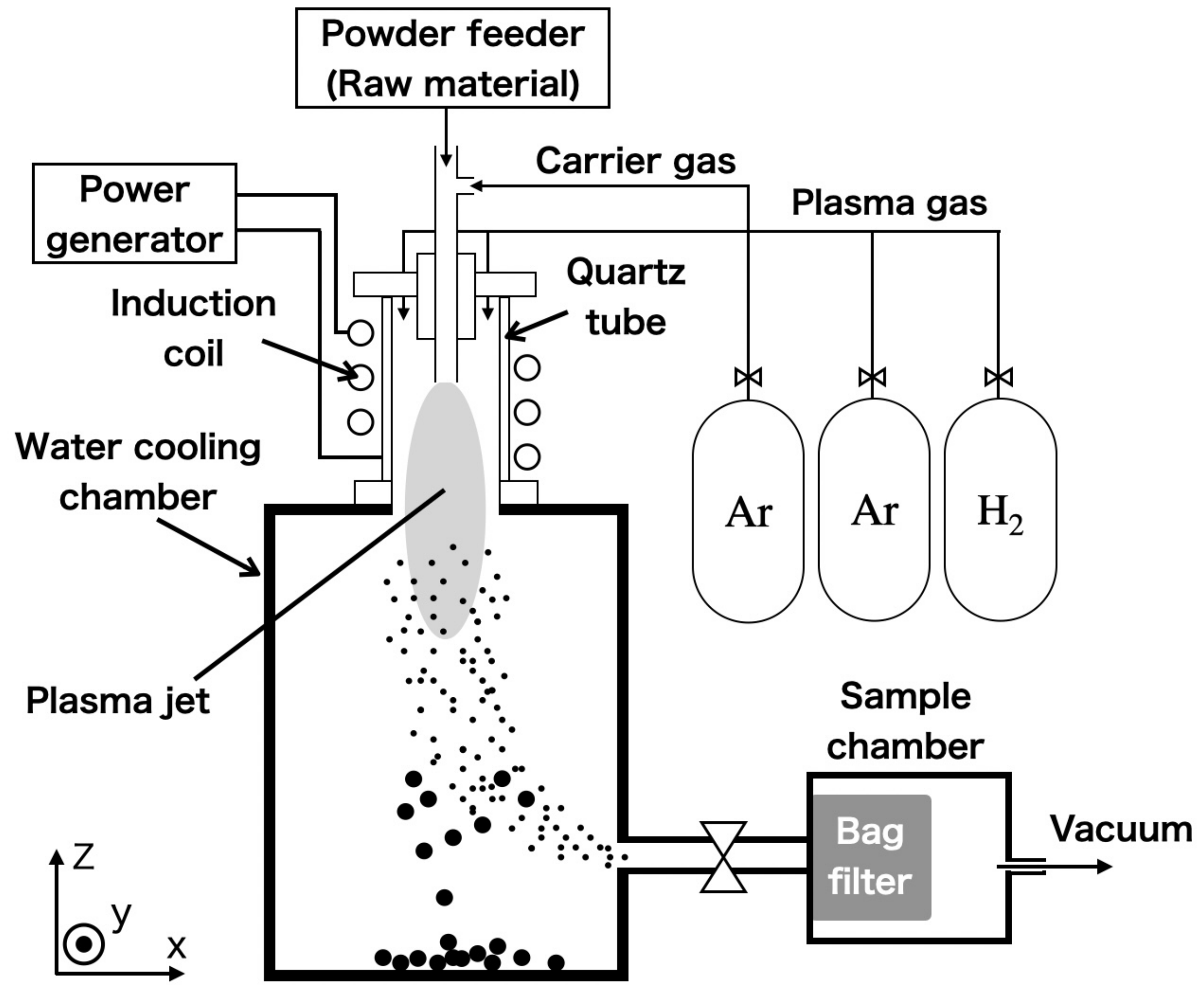}
\caption{Schematic drawing of nanoparticles production system with the RF thermal
plasma method.}
\label{fig:RFthemal}
\end{figure}

We also tested the Radio-Frequency (RF) thermal plasma method to produce vanadium nanoparticles and V-Ni nanoparticles. This method is categorized as a bottom-up approach. This method allows for the production of any alloy nanoparticle by adjusting the mixing ratio of different elemental raw powders. To produce high-purity nanoparticles and suppress oxidation, we used high-purity fine raw powders.\\
We used a test machine for the RF thermal plasma of Nissin Engineering Co., Ltd. As shown in Fig.~\ref{fig:RFthemal}, the entire apparatus consists of an RF power supply, a plasma torch, a reactor, and a sample compensation filter. The plasma torch consists of a solenoid coil (induction coil) surrounded with an insulating water-cooled quartz tube. 
The raw powder was introduced into the powder feeder along with $\mathrm{Ar}$ carrier gas. The thermal plasma (vapor of the substance) generated by the radio-frequency discharge was cooled by flowing it into a water-cooled stainless steel chamber. The resulting nano-sized powder was collected using a bag filter. \\
The raw materials fed into the powder feeder were natural vanadium fine powder and natural nickel fine powder provided by different manufacturers: $\mathrm{V(raw)_{1st}}$ fine powder (purity: $\mathrm{>99\,\%}$, particle size $\mathrm{\approx75\,\mu m}$), $\mathrm{V(raw)_{2nd}}$ fine powder (purity: $\mathrm{>99.7\,\%}$, particle size $\mathrm{\approx150\,\mu m}$), and $\mathrm{Ni(raw})$ fine powder (purity: $\mathrm{>99.99\,\%}$, particle size $\mathrm{\approx10\,\mu m}$).Vanadium nanoparticles and V-Ni nanoparticles were produced using $\mathrm{V(raw)_{1st}}$ and $\mathrm{V(raw)_{2nd}}$, respectively. \\
The sample preparation conditions are shown in Table~\ref{Table:RF_test}, where the input raw material amount is $m_\mathrm{in}$, 
the Ni raw material powder ratio is $k_\mathrm{Ni}$, and the yield is $m_\mathrm{col}$. 
The samples indicated by the subscripts $\mathrm{RF1}$ and $\mathrm{RF2}$ were prepared using $\mathrm{V_{1st}}$ fine powder and $\mathrm{V(raw)_{2nd}}$ fine powder as raw materials.
\begin{table}
\centering
\normalfont
\renewcommand{\arraystretch}{1.15}
\begin{tabular}{l lccc}
\toprule
\multicolumn{1}{l}{Sample} & \multicolumn{1}{c}{Raw powder} & \multicolumn{1}{c}{$m_{\mathrm{in}}\,(\mathrm{g})$} & \multicolumn{1}{c}{$k_\mathrm{{Ni}}\,(\mathrm{wt\%})$} & \multicolumn{1}{c}{$m_{\mathrm{col}}\,(\mathrm{g})$}\\\midrule
$\mathrm{V_{RF1}}$ & $\mathrm{V(raw)_{1st}}$ & \begin{tabular}[c]{@{}l@{}}  100\end{tabular} & -- & 3.2\\
$\mathrm{VNi_{RF1}}$ & $\mathrm{V(raw)_{1st}+Ni(raw)}$ & \begin{tabular}[c]{@{}l@{}} 50\end{tabular} & \begin{tabular}[c]{@{}l@{}}1.0\end{tabular} & 1.8\\
$\mathrm{V_{RF2}}$ & $\mathrm{V(raw)_{2nd}}$  & \begin{tabular}[c]{@{}l@{}} 500\end{tabular} & -- & 21.6\\
$\mathrm{VNi_{RF2}}$ & $\mathrm{V(raw)_{2nd}+Ni(raw)}$ & \begin{tabular}[c]{@{}l@{}} 250\end{tabular} & \begin{tabular}[c]{@{}l@{}}1.9\end{tabular} & 7.5\\
\bottomrule
\end{tabular}
    \caption{Test conditions using RF thermal plasma method. ($m_\mathrm{in}$: the input raw material amount, $k_\mathrm{Ni}$: the Ni raw material powder ratio, and $m_\mathrm{col}$: yield. Samples with subscripts $\mathrm{RF1}$ and $\mathrm{RF2}$: $\mathrm{V(raw)_{1st}}$ fine powder and $\mathrm{V(raw)_{2nd}}$ fine powder were used as raw materials, respectively. VNi sample: Ni(raw) powder mixed vanadium raw material powder was used.)}
    \label{Table:RF_test}
\end{table}
\vspace{1em}
The $\mathrm{VNi_{RF1}}$ and $\mathrm{VNi_{RF2}}$ sample powders were prepared by mixing vanadium raw powder and $\mathrm{Ni(raw)}$ raw powder, and the nickel contents were adjusted to $\mathrm{5.8\,wt\%}$ and $\mathrm{1.9\,wt\%}$, respectively.
 The mixing ratio of the raw powders used to produce V-Ni nanoparticles was determined by analyzing the nickel content of the prepared sample powders, taking into account the differences in the heat of vaporization and particle size of the raw materials. The nickel content of the $\mathrm{VNi_{RF2}}$ sample was adjusted to a lower level to prevent oxygen contamination.\\
The raw powder was placed in a sealed powder feeder, connected to a plasma torch, and fed at a rate of $\mathrm{110\sim120\,g/h}$. The plasma gas used was a mixture of Ar gas and a few percent of $\mathrm{H_2}$ as an oxygen reducing agent. The raw powder preparation and the collection of the sample powder were carried out in an $\mathrm{Ar}$ gas atmosphere using a glove box, and the oxygen concentration during the work was about $\mathrm{100\,ppm}$.

\subsection{Particle shape and main component}
\subsubsection{Observation of particle shape using SEM}
\begin{figure*}
\centering
\normalfont
     \includegraphics[width=0.3\columnwidth]{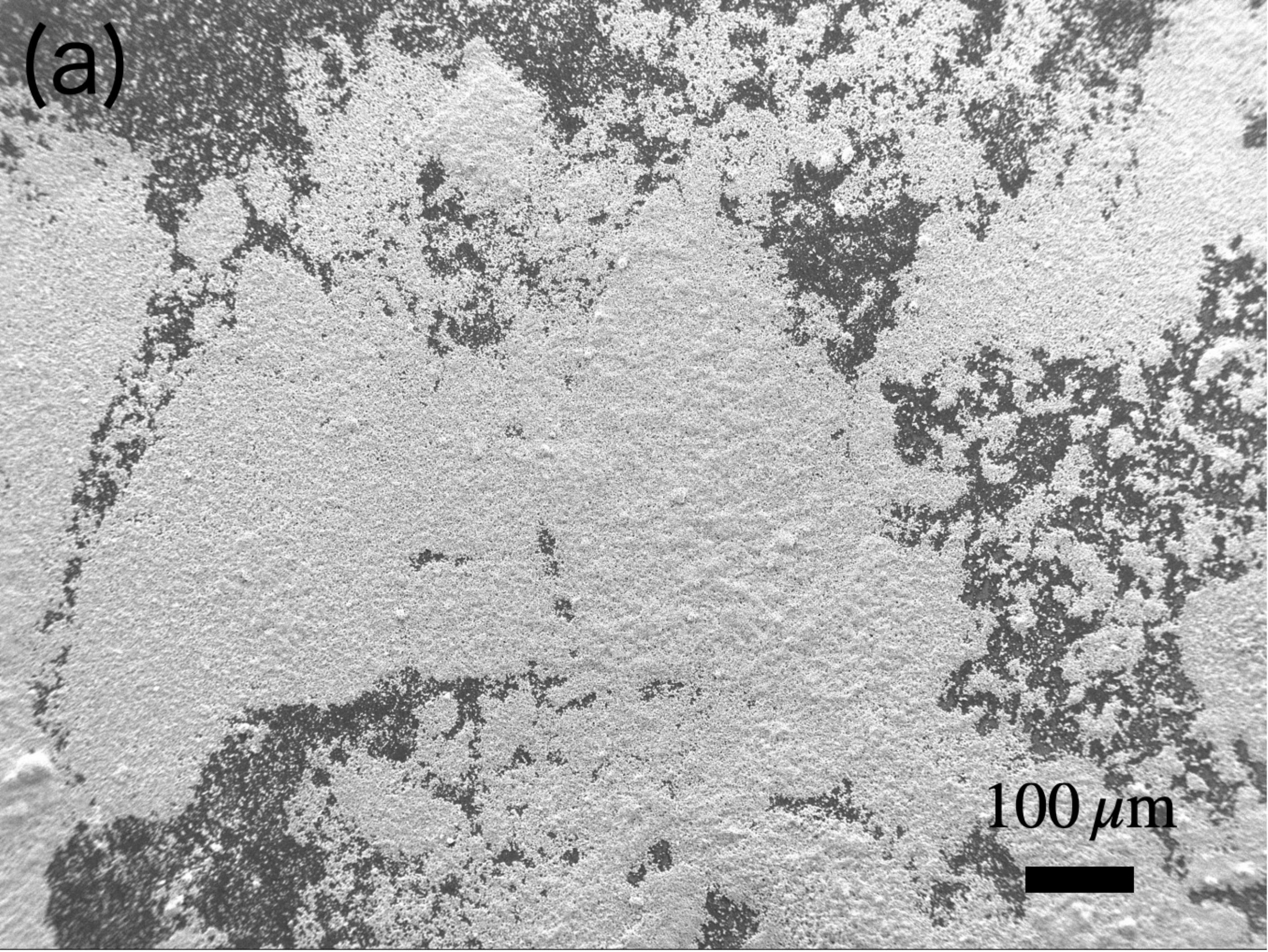}
     \includegraphics[width=0.3\columnwidth]{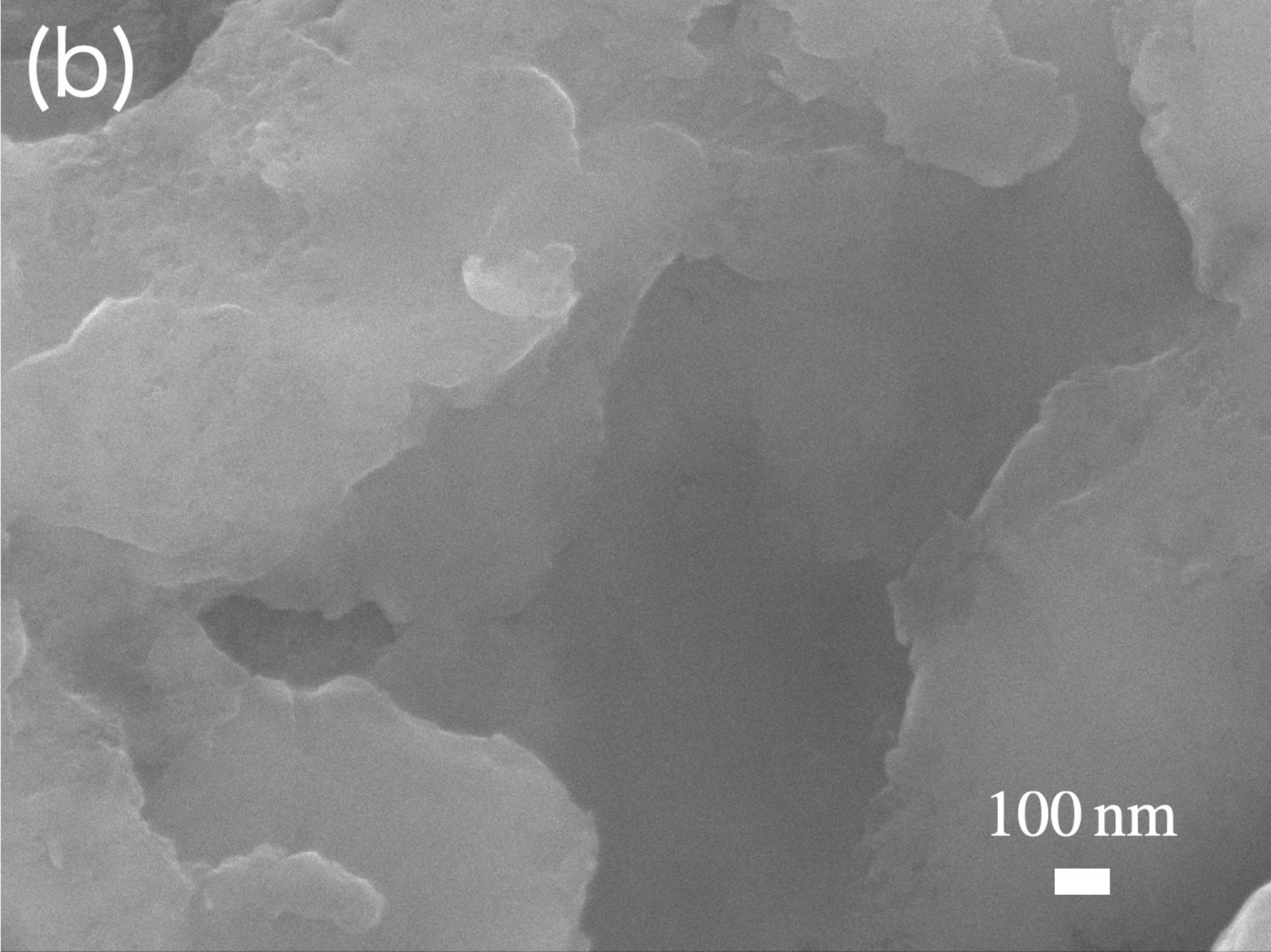}
     \includegraphics[width=0.3\columnwidth]{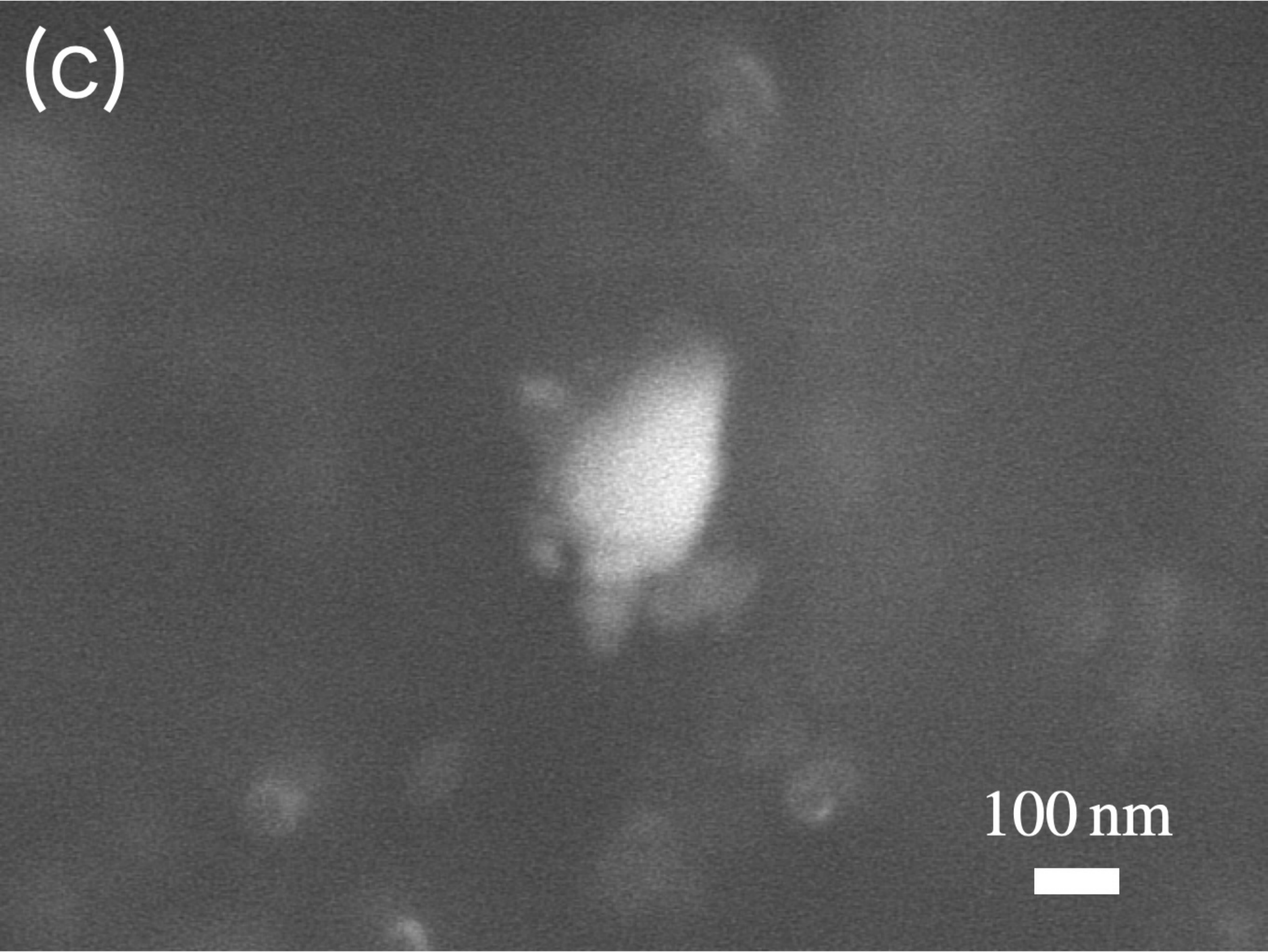}
     \includegraphics[width=0.3\columnwidth]{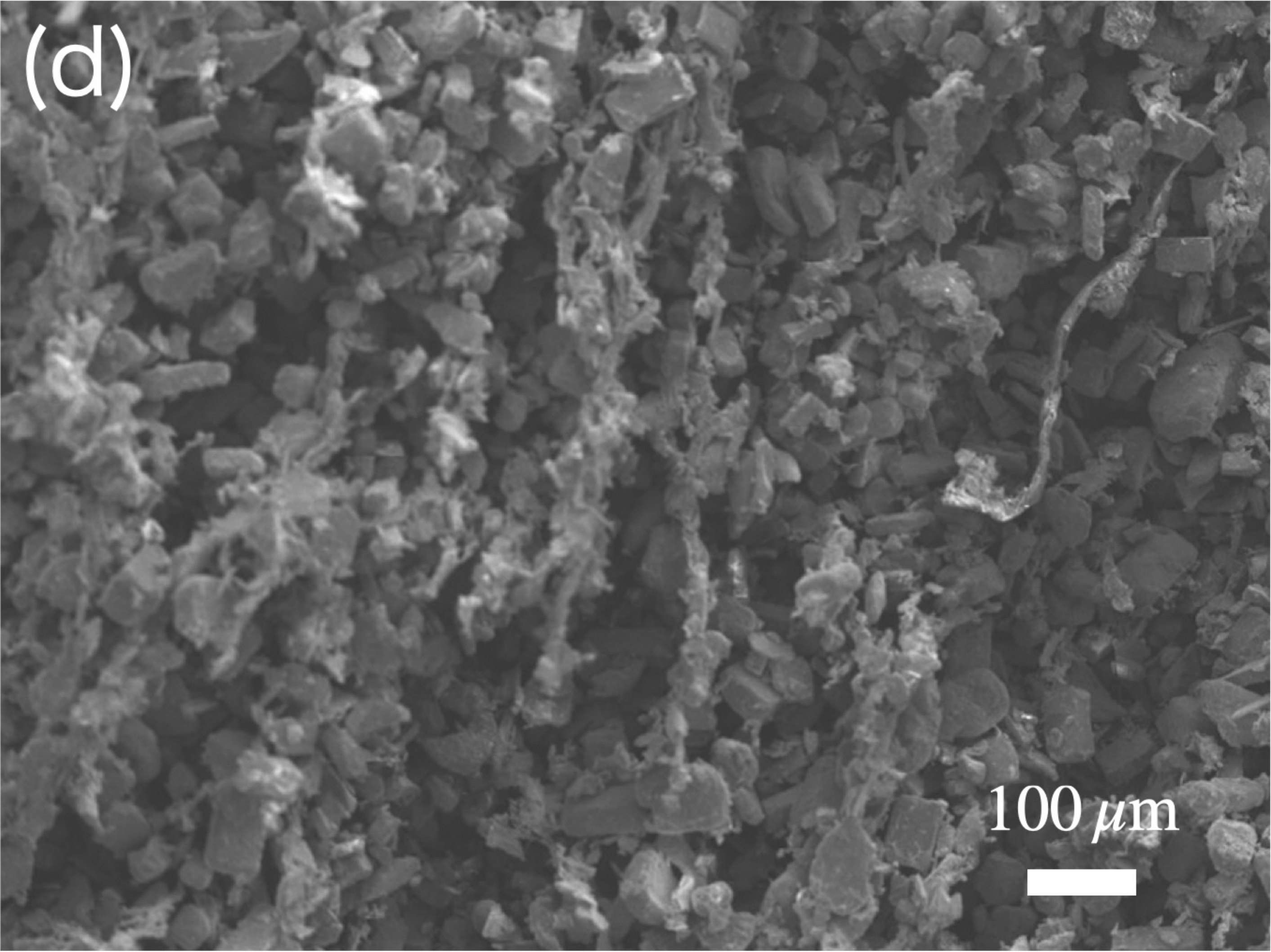}
     \includegraphics[width=0.3\columnwidth]{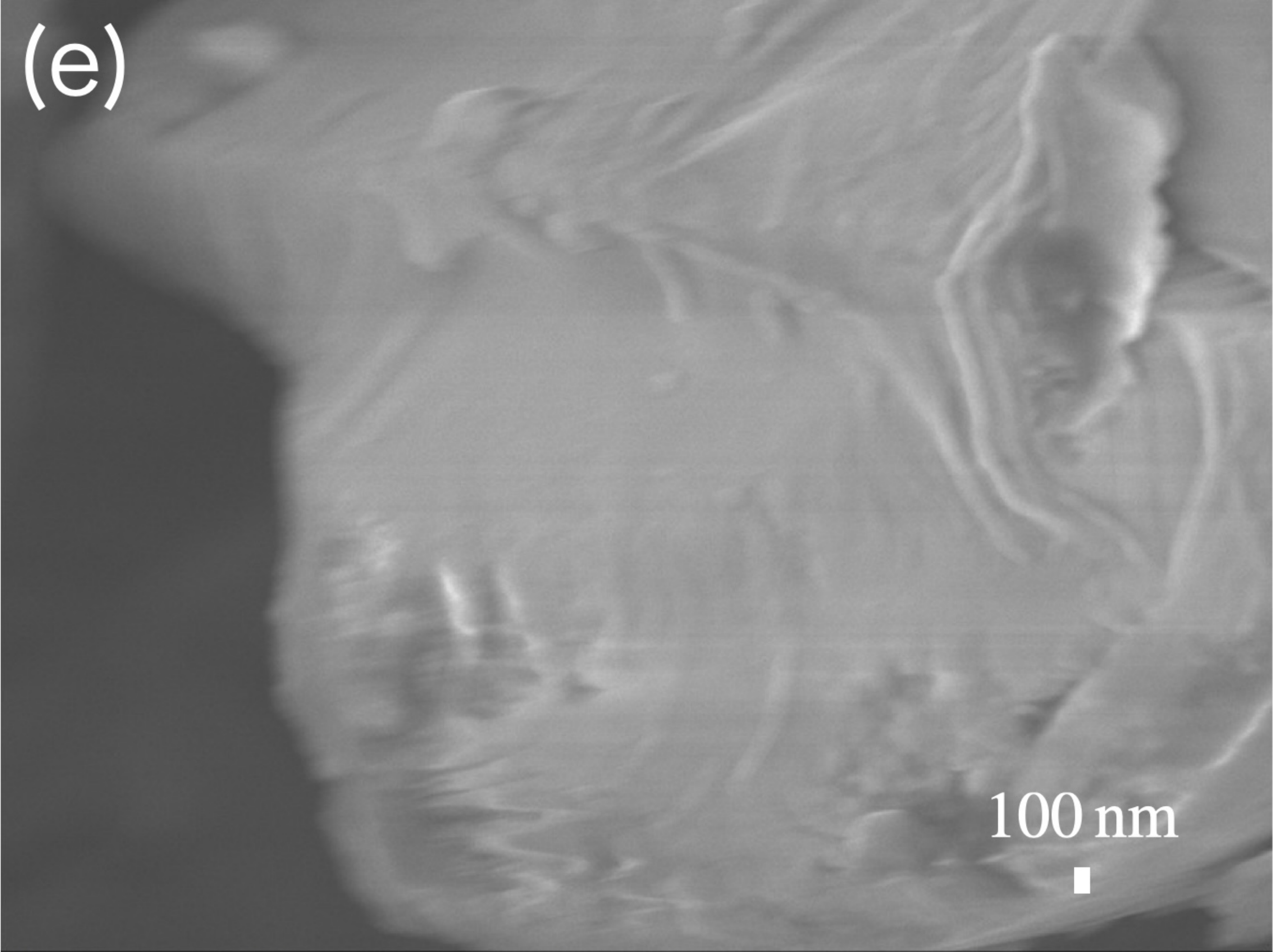}
     \includegraphics[width=0.3\columnwidth]{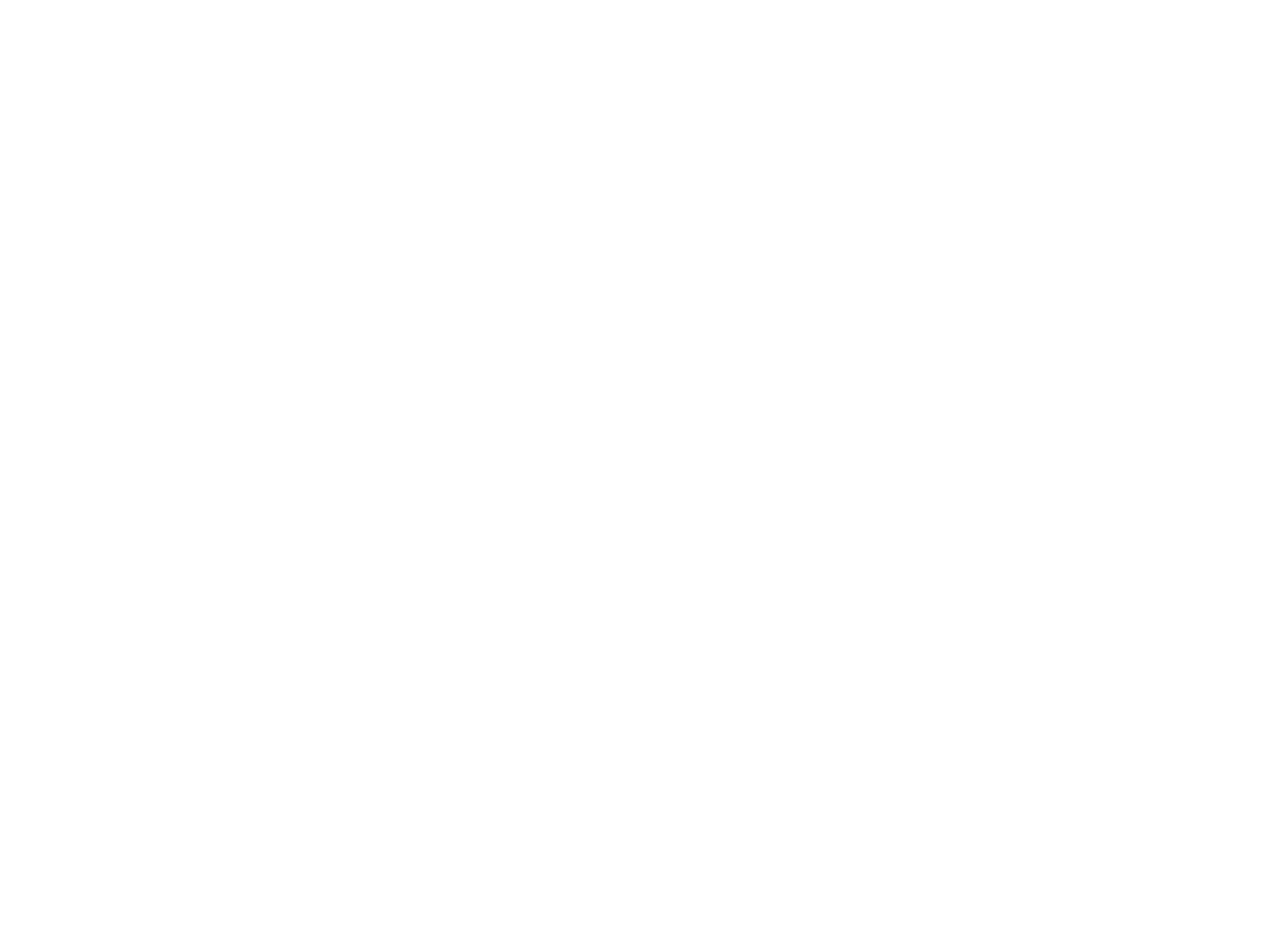}
     \includegraphics[width=0.3\columnwidth]{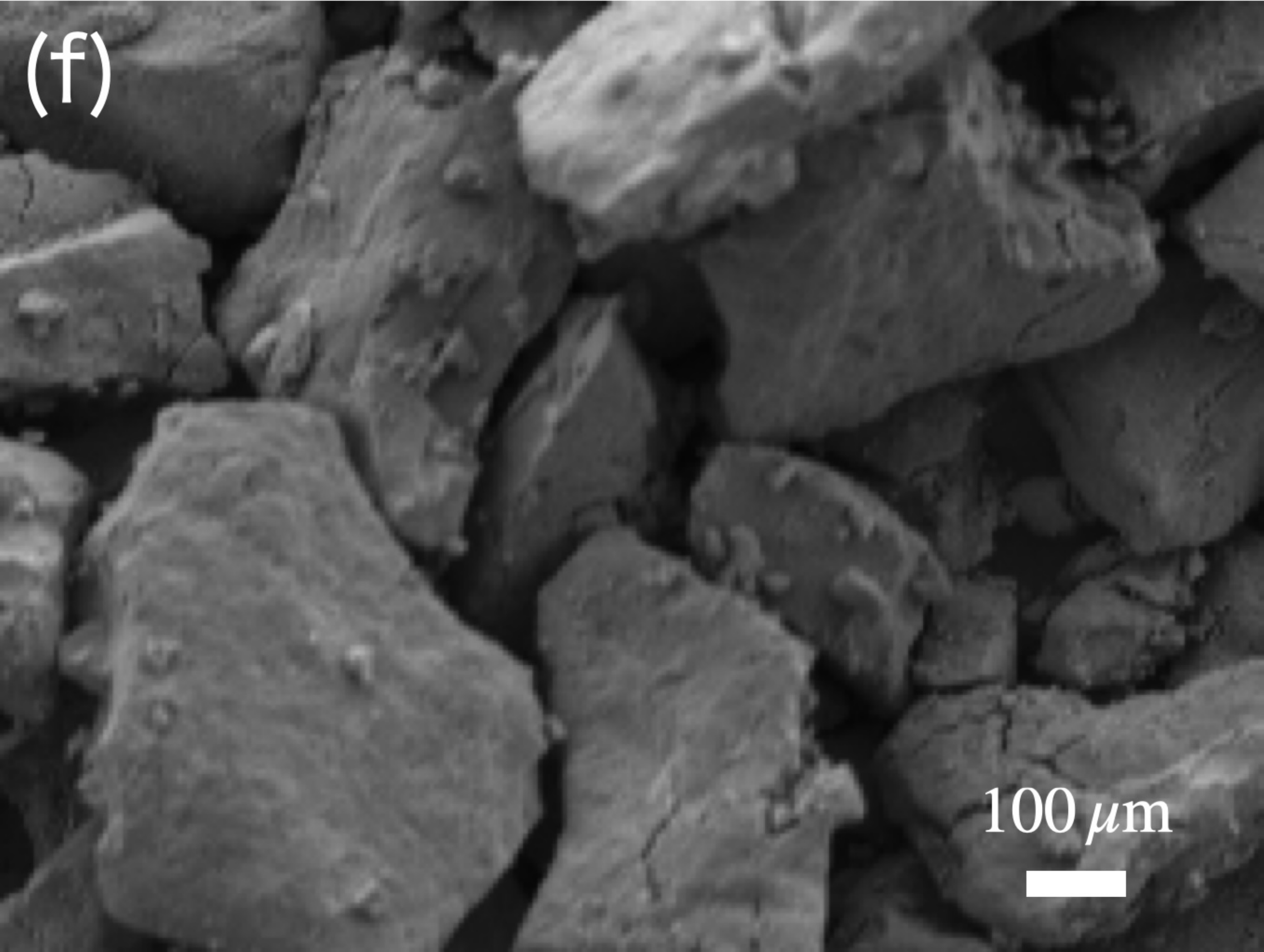}
     \includegraphics[width=0.3\columnwidth]{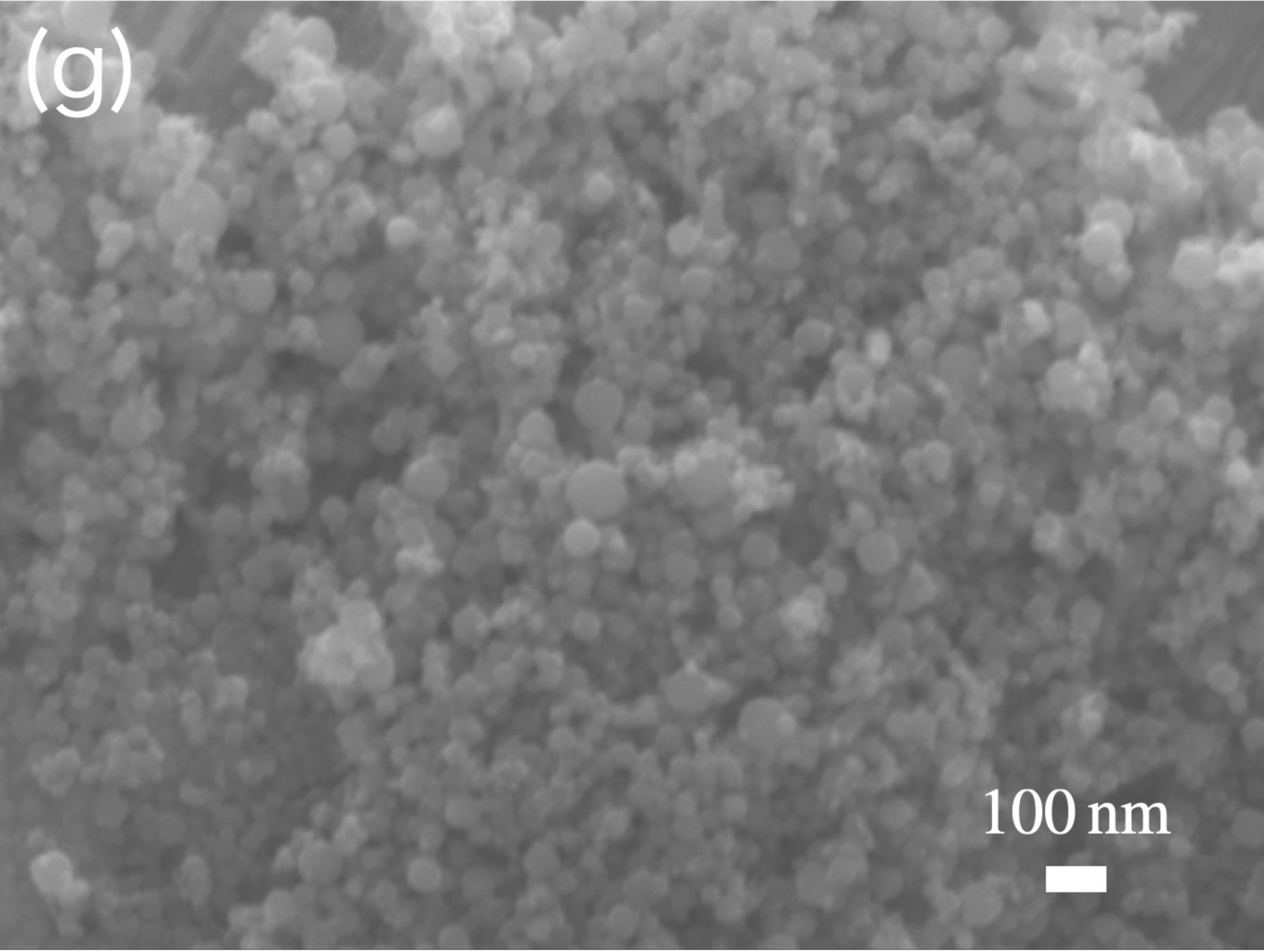}
     \includegraphics[width=0.3\columnwidth]{figs/SEM/Figure_7blank.pdf}
\caption{SEM images of three powder samples observed at different magnifications. (a) and (b) $\mathrm{VNi_{jm}}$ sample powder created by jet milling [magnification:$\mathrm{43.0\times}$ and $\mathrm{55.0k\times}$], (a) and (b) $\mathrm{VNi_{jm}}$ sample powder created by jet milling [magnification:$\mathrm{43.0\times}$ and $\mathrm{55.0k\times}$], (c) $\mathrm{V_{jm}}$ sample powder created by jet milling [magnification:$\mathrm{85.0k\times}$], (d) and (e) Commercially available vanadium nanopowder $\mathrm{V_{c.a.}}$ [magnification:$\mathrm{110\times}$ and $\mathrm{14.0k\times}$], (f) and (g) $\mathrm{V_{RF2}}$ sample powder created by RF thermal plasma [magnification:$\mathrm{95.0\times}$ and $\mathrm{60.0k\times}$]. Different scale bars are used to properly visualize the particle morphology at each magnification. ((a), (d) and (f)) Low-magnification SEM images, ((b), (c), (e) and (g))  High-magnification SEM images. }
\label{fig:SEM}
\end{figure*}

A field emission scanning electron microscope (FE-SEM) was used to check the shape and size of the prepared sample powder and commercially available vanadium nanopowder. For comparison with the sample prepared in this study, commercially available vanadium nanopowder (hereinafter referred to as $\mathrm{V_{c.a.}}$) was prepared, with nominal values of $99\,\%$ purity and particle size of 80$\sim$100$\,\mathrm{nm}$.\\
Figure~\ref{fig:SEM} shows SEM images of each sample observed at low and high magnification.
Figures~\ref{fig:SEM}(a), (b) and (c) show images of sample powders prepared using a jet mill at different magnifications. The $\mathrm{VNi_{jm}}$ sample powder, as shown in Figs.~\ref{fig:SEM}(a) and (b), was confirmed to have an irregular shape with diameters ranging from tens of $\mathrm{\mu m}$ to hundreds of $\mathrm{nm}$ was confirmed. Furthermore, Fig.~\ref{fig:SEM}(c) shows a high-magnification SEM image of a prototype sample (hereinafter referred to as $\mathrm{V_{jm}}$) prepared using natural vanadium plate (purity: $\mathrm{99.7\,\%}$) as the raw material. The jet milling process was performed similarly, and irregularly shaped particles with diameters of tens of micrometers were also observed, similar to $\mathrm{VNi_{jm}}$, but the presence of nanometer-scale particles was suggested.\\
Figures~\ref{fig:SEM}(d) and (e) show SEM images of the $\mathrm{V_{c.a.}}$ powder at different magnifications. Low-magnification observation reveals the presence of multiple irregularly shaped particles, approximately $\mathrm{\mu m}$ in size. High-magnification observation reveals the absence of boundaries between particles, suggesting a mixture of $\mathrm{\mu m}$-sized particles rather than aggregates of $\mathrm{nm}$-sized particles.\\
Sample powder prepared by RF thermal plasma is shown in Figs.~\ref{fig:SEM}(f) and (g) at different magnifications. The sample particles created using RF thermal plasma were spherical with diameters of less than $\mathrm{50\,nm}$, and each sample was confirmed to have a similar shape and size.\\
While the SEM image of $\mathrm{V_{jm}}$ suggested the presence of nanometer-scale particles, their number density was significantly lower than that of the RF thermal plasma-treated sample, and their particle morphology lacked uniformity. Therefore, the RF thermal plasma method provided nanoparticles more suitable for the requirements of small-angle scattering measurements. For this reason, further characterization of the RF thermal plasma-treated samples was carried out. In the following sections, we focus on the samples fabricated using the RF thermal plasma method.

\subsubsection{EDS analysis}
\begin{figure}[t]
\centering
    \centering
    \includegraphics[width=0.32\textwidth]{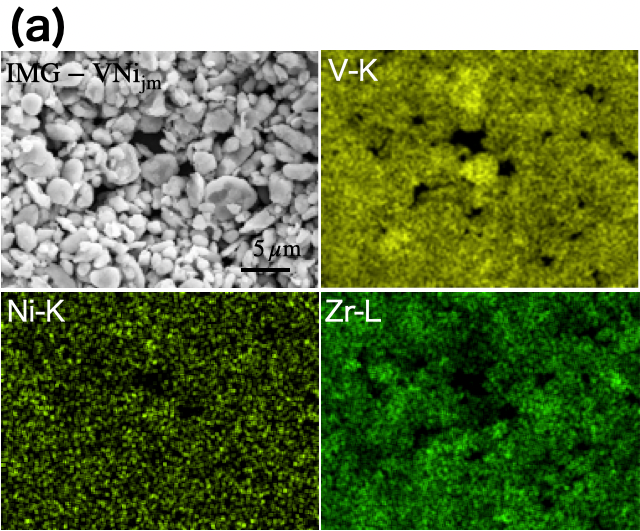}
    \includegraphics[width=0.32\textwidth]{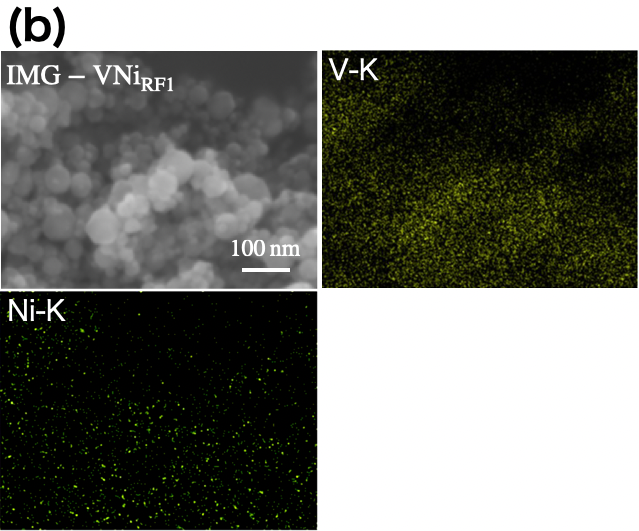}
    \includegraphics[width=0.32\textwidth]{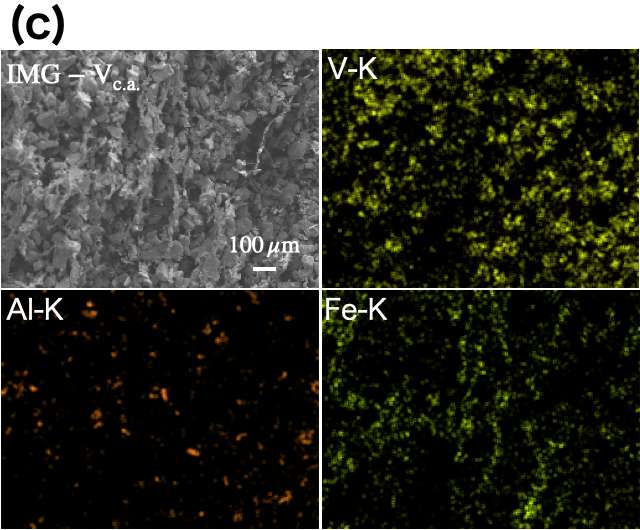}
\caption{Spatial distributions of each element in each sample powder measured with the EDS method. (a) $\mathrm{VNi_{jm}}$ sample powder [magnification:$\mathrm{\times}$], (b) $\mathrm{VNi_{RF1}}$ sample powder [magnification:$\mathrm{110\times}$], (c) $\mathrm{V_{c.a.}}$ powder [magnification:$\mathrm{110\times}$].}
\label{fig:MAP_sem}
\end{figure}

The metal element content ratios in each sample powder were analyzed using an Energy Dispersive X-ray Spectroscopy (EDS) mounted on the FE-SEM. The detected composition ratios for each sample are shown in Table~\ref{Table:SEM-EDS}, and the element distribution mapping results are shown in Fig.~\ref{fig:MAP_sem}. 

\begin{table*}[t]
\centering
\normalfont
\setlength{\tabcolsep}{0.2cm}
\renewcommand{\arraystretch}{1.2}
\begin{tabular}{l lllllll}
\toprule
 & \multicolumn{1}{c}{$\mathrm{VNi_{jm}}$} & \multicolumn{1}{c}{$\mathrm{V_{jm}}$} & \multicolumn{1}{c}{$\mathrm{V_{RF1}}$} & \multicolumn{1}{c}{$\mathrm{VNi_{RF1}}$} & \multicolumn{1}{c}{$\mathrm{V_{RF2}}$} & \multicolumn{1}{c}{$\mathrm{VNi_{RF2}}$} & \multicolumn{1}{c}{$\mathrm{V_{c.a.}}$} 

\\\midrule
Al  &1.44(5)    &\multicolumn{1}{c}{--}      &\multicolumn{1}{c}{--}    &\multicolumn{1}{c}{--} &\multicolumn{1}{c}{--} &\multicolumn{1}{c}{--} &5.06(21)\\
Si   &\multicolumn{1}{c}{--}   &0.35(3)  &\multicolumn{1}{c}{--}   &\multicolumn{1}{c}{--}   &\multicolumn{1}{c}{--}   &\multicolumn{1}{c}{--}   &\multicolumn{1}{c}{--}\\
V   &66.41(37)  &99.65(67)     &100.00(103)              &94.94(141) & 100.00(37) &98.41(110) &47.97(148)\\
Cr  &3.34(12)   &\multicolumn{1}{c}{--}      &\multicolumn{1}{c}{--}    &\multicolumn{1}{c}{--} &\multicolumn{1}{c}{--} &\multicolumn{1}{c}{--} &2.62(56)\\
Mn  &\multicolumn{1}{c}{--}   &\multicolumn{1}{c}{--}  &\multicolumn{1}{c}{--}    &\multicolumn{1}{c}{--} &\multicolumn{1}{c}{--} &\multicolumn{1}{c}{--} &4.47(70)\\
Fe  &\multicolumn{1}{c}{--}   &\multicolumn{1}{c}{--}  &\multicolumn{1}{c}{--}    &\multicolumn{1}{c}{--} &\multicolumn{1}{c}{--} &\multicolumn{1}{c}{--} &39.89(160)\\
Ni  &3.34(12)&\multicolumn{1}{c}{--}    &\multicolumn{1}{c}{--}    &5.06(112) &\multicolumn{1}{c}{--} &1.46(56) &\multicolumn{1}{c}{--}\\
Cu  &1.12(10)&\multicolumn{1}{c}{--}    &\multicolumn{1}{c}{--}    &\multicolumn{1}{c}{--} &\multicolumn{1}{c}{--} &\multicolumn{1}{c}{--} &\multicolumn{1}{c}{--}\\
Zn  &1.43(12)&\multicolumn{1}{c}{--}    &\multicolumn{1}{c}{--}    &\multicolumn{1}{c}{--} &\multicolumn{1}{c}{--} &\multicolumn{1}{c}{--} &\multicolumn{1}{c}{--}\\
Zr  &21.48(21)&\multicolumn{1}{c}{--}  &\multicolumn{1}{c}{--}    &\multicolumn{1}{c}{--} &\multicolumn{1}{c}{--} &\multicolumn{1}{c}{--} &\multicolumn{1}{c}{--}\\
\bottomrule
\end{tabular}
    \caption{Composition ratio of metal elements contained in each sample powder by EDS analysis in unit of $\mathrm{wt\%}$. Elements without a value in the table are below the detection limit.}
    \label{Table:SEM-EDS}
\end{table*}

In addition to the major elements, multiple metal elements were detected in the $\mathrm{VNi_{jm}}$ and $\mathrm{V_{c.a.}}$ sample powders. The main contaminant in the $\mathrm{VNi_{jm}}$ sample powder was $\mathrm{Zr}$. As shown in Fig.~\ref{fig:MAP_sem}(a), $\mathrm{Zr}$ was detected throughout the powder at approximately $\mathrm{20\,wt\%}$, similar to the $\mathrm{Ni}$ contained in the V-Ni alloy used as the raw material. $\mathrm{Fe}$, the main element in metal snips, was found to be present at approximately $\mathrm{150\,ppm}$ in the V-Ni alloy coarsely crushed before being fed into the jet mill. However, the Fe content in the $\mathrm{VNi_{jm}}$ sample powder was below the detection limit, likely due to surface material of jet mill equipment. $\mathrm{Fe}$ contamination, primarily at approximately $\mathrm{40\,wt\%}$, was detected in the $\mathrm{V_{c.a.}}$ sample powder, and was also detected in areas separate from the scattered vanadium, as shown in Fig.~\ref{fig:MAP_sem}(c). The coherent nuclear scattering lengths of $\mathrm{Zr}$ and Fe are $\mathrm{9.45(2)\,fm}$ and $\mathrm{7.16(3)\,fm}$, respectively. To achieve a coherent nuclear scattering length comparable to that of natural vanadium, the content must be kept below $\mathrm{15\,wt\%}$, even if no other elements are present. The $\mathrm{VNi_{jm}}$ and $\mathrm{V_{c.a.}}$ sample powders contained contamination levels exceeding the allowable range and did not meet the target requirements for neutron scattering experiments.  On the other hand, in the $\mathrm{V_{jm}}$ sample powder and the sample prepared using RF thermal plasma, only the major elements V or Ni were detected, while other metallic elements were below the EDS detection limit. The difference between the contamination levels of the $\mathrm{V_{jm}}$ and $\mathrm{VNi_{jm}}$ samples produced with the jet mill method may be not only due to differences in raw material size and grinding time, but also due to differences in material hardness and viscosity. 
Increasing the grinding time could improve the efficiency of nano-sized particle production, but that also increases the risk of contamination, which may be resolved with the future development. As shown in Fig.~\ref{fig:MAP_sem}(b), in the sample prepared by mixing Ni with the raw material fine powder using the RF thermal plasma method, Ni was detected in the same locations as vanadium, similar to the V-Ni alloy, and the uniform distribution indicated alloying at the atomic level. The nickel contents of the $\mathrm{VNi_{RF1}}$ and $\mathrm{VNi_{RF2}}$ sample powders, as a percentage of the total vanadium and nickel masses, were $\mathrm{5.06(112)\,wt\%}$ and $\mathrm{1.46(52)\,wt\%}$, respectively, which were within the target nickel content within the analytical error of EDS. This indicates that the RF thermal plasma method can achieve a blending ratio control accuracy of approximately $\pm \mathrm{0.5\,wt\%}$, suggesting that this method can be used to fabricate alloy nanoparticles with the desired blending ratio. Under current manufacturing conditions, targets created by the RF thermal plasma method possessed particle sizes suitable for small-angle scattering targets, and metal contamination was within acceptable limits. Therefore, more detailed analytical results of these samples using oxygen content and trace element analysis are described in Sec.~\ref{sec:3.3} and Sec.~\ref{sec:4.1}.

\subsection{Oxygen concentration}
\label{sec:3.3}
To measure the oxygen content of the sample powders produced using RF thermal plasma, non-dispersive infrared absorption (NDIR) analysis was performed without exposure to air. The sample analysis results are shown in Table~\ref{Table:EMGA_sample}. Measurements were performed three times for each sample, and the averages of the measured values were used to calculate total coherent scattering lengths, with the standard deviations as the errors. To avoid exposing the sample powders to air, all sample processings up to analysis were performed in a glove box filled with $\mathrm{Ar}$ gas.\\
Table~\ref{Table:EMGA_sample} shows that the sample produced using $\mathrm{V(raw)_{2nd}}$ as the raw material of the RF thermal plasma method had a lower oxygen content than the sample produced using $\mathrm{V(raw)_{1st}}$. Among the samples in Table~\ref{Table:EMGA_sample}, the $\mathrm{V_{RF2}}$ sample powder had the lowest oxygen content of $\mathrm{6.15(14)\,wt\%}$. While the NDIR analysis provides only the ensemble-averaged oxygen concentration, significant spatial inhomogeneity would modify the q-dependent scattering profile. As will be discussed in Sec.~\ref{sec:4.2}, the scattering profiles analyzed in this work are adequately reproduced without introducing additional structural features within the present experimental precision. The oxygen contamination in $\mathrm{VNi_{RF1}}$ and $\mathrm{VNi_{RF2}}$ sample powders were expected to be less than $\mathrm{3\,wt\%}$, but more than $\mathrm{10\,wt\%}$ of oxygen was detected in both samples produced with them.

\begin{table}[h]
\centering
\normalfont
\setlength{\tabcolsep}{0.28cm}
\renewcommand{\arraystretch}{0.9}
\begin{tabular}{c c c c c}
\midrule
\multicolumn{1}{c}{} & \multicolumn{1}{c}{$\mathrm{V_{RF1}}$} & \multicolumn{1}{c}{$\mathrm{VNi_{RF1}}$} & \multicolumn{1}{c}{$\mathrm{V_{RF2}}$} & \multicolumn{1}{c}{$\mathrm{VNi_{RF2}}$}\\\midrule
Oxygen & 14.6(3) & 13.4(3) & 6.15(14) & 11.4(1)\\
\midrule
\end{tabular}
    \caption{Quantitative analysis of Oxygen in the each sample powder in unit of $\mathrm{wt\%}$.}
    \label{Table:EMGA_sample}
\end{table}
Table~\ref{Table:ICP_Raw} shows the composition ratio of the raw fine powder for RF thermal plasma. The results for $\mathrm{V(raw)_{1st}}$ and $\mathrm{Ni(raw)}$ are measured by inductively coupled plasma atomic emission spectroscopy (ICP-AES). The sample for ICP-AES measurement was produced by resolving a weighed sample powder into $\mathrm{8\,ml}$ of aqua regia. The sample solution was heated at $230\mathrm{\,^{\circ}C}$ for 20 minutes using a microwave device to promote decomposition, and then heated at $230\mathrm{\,^{\circ}C}$ for another 20 minutes. The volume was then adjusted to a constant value using a $\mathrm{50\,ml}$ volumetric flask, and the solution was diluted 200 times with water for measurement. The results for the $\mathrm{V(raw)_{2nd}}$ fine powder correspond to the oxygen concentration at the time of shipment. 
"Unidentifiable" in the table refers to the sum total of light elements that cannot be detected by ICP-AES, such as moisture and organic matter adhering to the sample surface, oxygen in metal oxides, and carbon, and the total concentrations of unidentifiable contaminants were estimated as $\mathrm{3.2\,wt\%}$ and $\mathrm{0.9\,wt\%}$ for the $\mathrm{V(raw)_{1st}}$ and the $\mathrm{Ni(raw)}$ samples, respectively. This difference in the concentrations of the unidentifiable contaminants can be explained if the unidentifiable contaminants are dominated by oxygen, since it is known that vanadium metal is more easily oxidized than nickel. The oxygen content of $\mathrm{V(raw)_{2nd}}$ fine powder is $\mathrm{0.186\,wt\%}$, and as shown in Table~\ref{Table:ICP_Raw}, the sample made using $\mathrm{V(raw)_{2nd}}$ as the raw material for RF thermal plasma has a lower oxygen content than the sample made using $\mathrm{V(raw)_{1st}}$ as the raw material.
\begin{table}[t]
\centering
\normalfont
\setlength{\tabcolsep}{0.4cm}
\renewcommand{\arraystretch}{0.9}
\begin{tabular}{llll}
\toprule\toprule
Element & $\mathrm{V(raw)_{1st}}$ & $\mathrm{V(raw)_{2nd}}$ & $\mathrm{Ni(raw)}$\\\midrule
C  & \multicolumn{1}{c}{--} & 0.0001 & \multicolumn{1}{c}{--}\\
O  & \multicolumn{1}{c}{--} & 0.186  & \multicolumn{1}{c}{--}\\
Mg & 0.0004 & \multicolumn{1}{c}{--} & \multicolumn{1}{c}{--}\\
Al & 0.005 & 0.001 & 0.0003\\
Si & 0.034 & 0.032 & 0.004\\
Ca & 0.005  & \multicolumn{1}{c}{--} & 0.005\\
Ti & 0.001  & \multicolumn{1}{c}{--} & \multicolumn{1}{c}{--}\\
V  & 96.8   & \multicolumn{1}{c}{Bal.} & 0.01\\
Fe & \multicolumn{1}{c}{--} & 0.008 & \multicolumn{1}{c}{--}\\
Ni & 0.002  & \multicolumn{1}{c}{--} & 99.1\\
Mo & \multicolumn{1}{c}{--} & 0.005 & \multicolumn{1}{c}{--}\\\midrule
Total & 96.8 & \multicolumn{1}{c}{--} & 99.1\\
Unidentifiable & 3.2 & \multicolumn{1}{c}{--} & 0.9\\
\bottomrule\bottomrule
\end{tabular}
    \caption{Elemental compositions of raw materials used in the RF thermal plasma method in units of $\mathrm{wt.\,\%}$. Elements with no values in the table are either below the detection limit or contain balance.}
    \label{Table:ICP_Raw}
\end{table}\vspace{1em}

\section{Nuclear scattering cross section}
\label{sec:4}
\subsection{Calculation of neutron scattering length for nanoparticle powders}
\label{sec:4.1}
\begin{table*}
\centering
\normalfont
\setlength{\tabcolsep}{0.5cm}
\renewcommand{\arraystretch}{1.0}
\begin{tabular}{l llllll}
\toprule\toprule
\multicolumn{1}{l}{} & \multicolumn{1}{c}{$k\mathrm{\,(wt\%)}$} & \multicolumn{1}{c}{$b_\mathrm{coh}\,(\text{fm})$} & \multicolumn{1}{c}{$\sigma_\mathrm{inc}\,(\mathrm{b})$} & \multicolumn{1}{c}{$\sigma_\mathrm{abs}\,(\mathrm{b})$}\\\midrule
C   &0.67(1)                & 6.6484(13)   &0.001(4)    &0.00350(7)\\
O   &6.15(14)               & 5.805(4)     &0.0000(8)   &0.00019(2)\\
B   &0.0018(28)             & 5.30(4)      &1.70(12)    &767(8)\\
Mg  &0.003(9)               & 5.375(4)     &0.08(6)     &0.063(3)\\
Al  &0.0088(1)              & 3.449(5)     &0.0082(6)   &0.231(3)\\
Si  &0.043(1)               & 4.15071(22)  &0.004(8)    &0.171(3)\\
Ca  &0.0036(30)             & 4.70(2)      &0.05(3)     &0.43(2)\\
Ti  &0.0012(1)              & -3.730(13)   &2.87(3)     &6.09(13)\\
V   &92.96(1)               & -0.555(3)    &5.08(6)     &5.08(4)\\
Cr  &0.028(1)               & 3.635(7)     &1.83(2)     &3.05(6)\\
Mn  &0.002(1)               & -3.750(18)   &0.40(11)    &13.3(2)\\
Fe  &0.084(1)               & 9.45(2)      &0.40(11)    &2.56(3)\\
Zn  &0.009(1)               & 5.680(5)     &0.07(7)     &1.11(2)\\\midrule
$\mathrm{V_{RF2}}$ &100 & 0.719(23) & 4.08(2) &4.64(87)\\\bottomrule
\bottomrule
\end{tabular}
    \caption{Chemical concentrations and nuclear parameters of the detected elements.}
    \label{Table:SL_RF2nd}
\end{table*}
The coherent nuclear scattering length and neutron cross section of the $\mathrm{V_{RF2}}$ sample powder were calculated from the composition ratios obtained by trace element analysis. The data list and composition ratios of the coherent nuclear scattering length $b_\mathrm{coh}$, incoherent nuclear scattering cross section $\sigma_\mathrm{inc}$, and absorption cross section $\sigma_\mathrm{abs}$ for each detected element are shown in Table~\ref{Table:SL_RF2nd}. The neutron scattering length and cross section data list used for the calculations was the data published by NIST~\cite{Sears01011992} and the latest V coherent nucleus scattering length data~\cite{PhysRevLett.132.023402}.
The neutron cross section values  of the $\mathrm{V_{RF2}}$ sample powders weighted by the composition ratio are shown in the last row of Table~\ref{Table:SL_RF2nd}.
The neutron cross sections of $\mathrm{V_{RF2}}$ were calculated using the mean coherent nuclear scattering length $\langle b_\mathrm{coh}\rangle=\sum_{i}k_ib_\mathrm{coh,i}$ and the absorption and incoherent scattering cross sections $\langle \sigma\rangle=k_i\sigma_i$, where $k_i$ represents the abundance of the $i$-th element detected from $\mathrm{V_{RF2}}$. The error in the calculated values was estimated from the measurement error in the composition analysis and the error in the neutron scattering length and cross section data list. Trace element analysis was performed using ICP-AES, with the same pretreatment conditions as in Sec.~\ref{sec:3.3}. Measurement error was calculated from the standard deviation of repeated measurements, as with infrared absorption analysis. The major elements $\mathrm{V}$ and $\mathrm{Ni}$ were measured three times using samples of approximately $\mathrm{10\,mg}$ for each. Trace metal elements were measured three times using samples of approximately $\mathrm{200\,mg}$ for each powder. The concentration of carbon in the sample powder was measured using SEM-EDS. In order to avoid the influence of the carbon atoms contained in the backing foil, a gold foil was used for backing, and the sample powder was attached on it. The composition ratios were calculated by correcting the composition ratios of other elements based on the oxygen content of the sample shown in Table~\ref{Table:EMGA_sample}, and then renormalizing the heavy element and carbon contents to a total of $\mathrm{100\,wt\%}$.  The average coherent nuclear scattering length of the $\mathrm{V_{RF2}}$ sample particles was calculated to be $\langle b_\mathrm{coh} \rangle=\mathrm{0.719(23)\,fm}$, whose absolute value is about $\mathrm{30\,\%}$ larger than that of natural vanadium. 
The error in the estimated $b_\mathrm{coh}$ is about $3\,\mathrm{\%}$, which is mainly due to the error in oxygen content of $0.14\,\mathrm{wt\%}$.

\subsection{Measurement of particle size distribution}
\label{sec:4.2}
\begin{figure}
    \centering
    \includegraphics[width=0.4\textwidth]{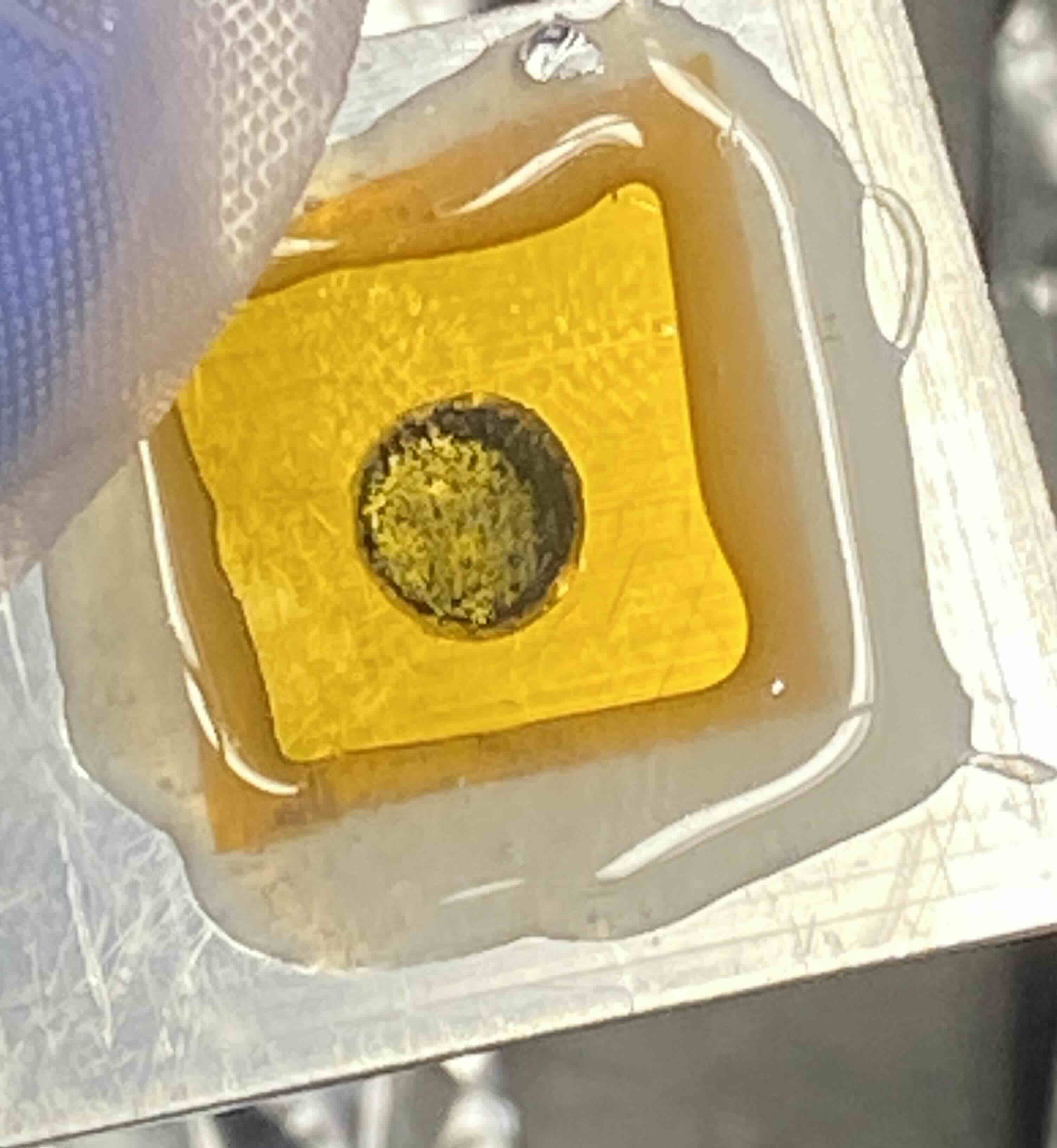}
    \caption{$\mathrm{V_{RF2}}$ sample powder attached to Kapton tape for SAXS measurements.}
    \label{fig:SAXS_sample}
\end{figure}

\begin{figure*}
    \centering
    \includegraphics[width=0.9\textwidth]{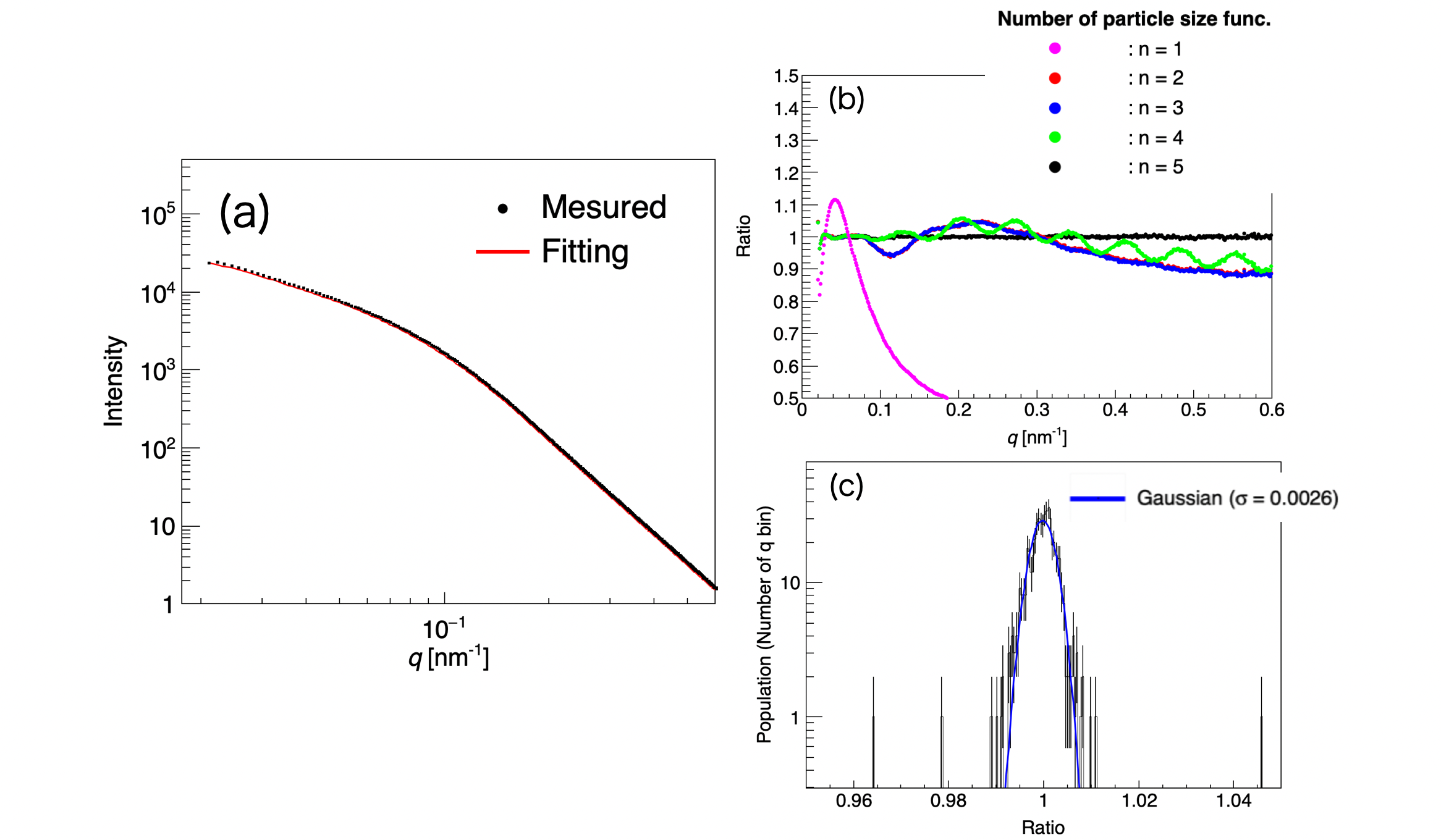}
    \caption{Small-angle X-ray scattering (SAXS) data measured at BL8S3 in Aichi-SR. (a) $q$ distribution and fit line obtained from the $\mathrm{V_{RF2}}$ sample powder target (b) Ratio of SAXS data to theoretical curve when particle size distribution function is increased. (c) Distribution of difference between experimental spectrum and the fitting function as a function of $q$.}
    \label{fig:SAXS_q}
\end{figure*}

The particle size distribution of the $\mathrm{V_{RF2}}$ sample powder was obtained by small-angle X-ray scattering (SAXS) measurements performed at BL8S3~\cite{Sugiyama_20170629} at the Aichi Synchrotron Radiation Center (Aichi-SR). The X-ray energy was $\mathrm{8.2\,KeV}$. The $\mathrm{V_{RF2}}$ sample powder was held in a hole with a diameter of $\mathrm{10\,mm}$ in an aluminum plate, covered with 13 $\mathrm{\mu m}$-thick Kapton tapes. Figure~\ref{fig:SAXS_sample} shows the picture of the sample assembly.
The sample was prepared in a vacuum glove box filled with $\mathrm{Ar}$ gas, and the corners of the attached Kapton tape were solidified with Araldite. An empty cell was also created in the same way. The measured SAXS data for the $\mathrm{V_{RF2}}$ sample powder are shown in Fig.~\ref{fig:SAXS_q}.
Figure~\ref{fig:SAXS_q}(a) shows the intensity at the momentum transfer ($q=4\pi\sin\theta/\lambda$) per unit solid angle for the $\mathrm{V_{RF2}}$. The SAXS profile was reproduced assuming a hard-sphere form factor combined with a superposition of size-distribution functions. Within the present experimental precision, no additional structural parameters were required to describe the data. The following function was used to fit the experimental spectra and obtain particle size and shape functions:
\begin{equation}
\begin{aligned}
P_{\mathrm{sum}}(q)
&= \sum_{k=1}^{n} P_k(q) \\
&\propto A \int 
\left[\sum_{k=1}^{n} a_k f_k(R,\bar{R}_k,\sigma_k)
\right]
\frac{F^2(q,R)}{V(R)}\, dR .
\end{aligned}
\label{SAXS_fitfunc}
\end{equation}

In Eq.(\ref{SAXS_fitfunc}), $f_n(R,\bar{R}_n,\sigma_n)$ and $F(q,R)$ stand for the n-th radius-distirbution function of the particle and the form factor of a particle with radius $R$, respectively.  Here, a hard sphere model was used for the shape parameters, i.e., $F(q,R) = 3\rho V(R)(\sin{(qR)}-qR\cos{(qR)})/{(qR)^3}$), and $V(R)$ is the volume of a sphere of radius $R$.
The fitting parameters are the variance $\sigma_n$ included in the particle size distribution, the mean particle size $\bar{R}_n$, the scale factor $a_n$ for the n-th particle size distribution function, and the scale factor $A$ for the entire function. The particle size distribution is calculated as $\sum_n a_nf_n(R,\bar{R}_n,\sigma_n)$ and normalized to one. The scale factor $A$ is a parameter used to normalize the measured data to the scattering intensity and does not affect the $q$ dependence. Fit analysis was performed by increasing the particle size distribution function in Eq.(\ref{SAXS_fitfunc}) to reproduce the $q$ distribution of the SAXS data. This fitting analysis was performed under the condition that the particle radius $R$ was $\mathrm{50\,nm}$ or less. Fig.~\ref{fig:SAXS_q}(b) shows the ratio of the SAXS data to the fit curve, and it can be seen that the SAXS data can be reproduced when the number of functions is increased up to five. The accuracy of the fitting was examined by checking the distribution of the difference between the fitting function and the experimental spectrum. Figure~\ref{fig:SAXS_q}(c) shows the distribution of the difference at each value of $q$. 
\begin{figure}
    \centering
    \includegraphics[width=0.6\textwidth]{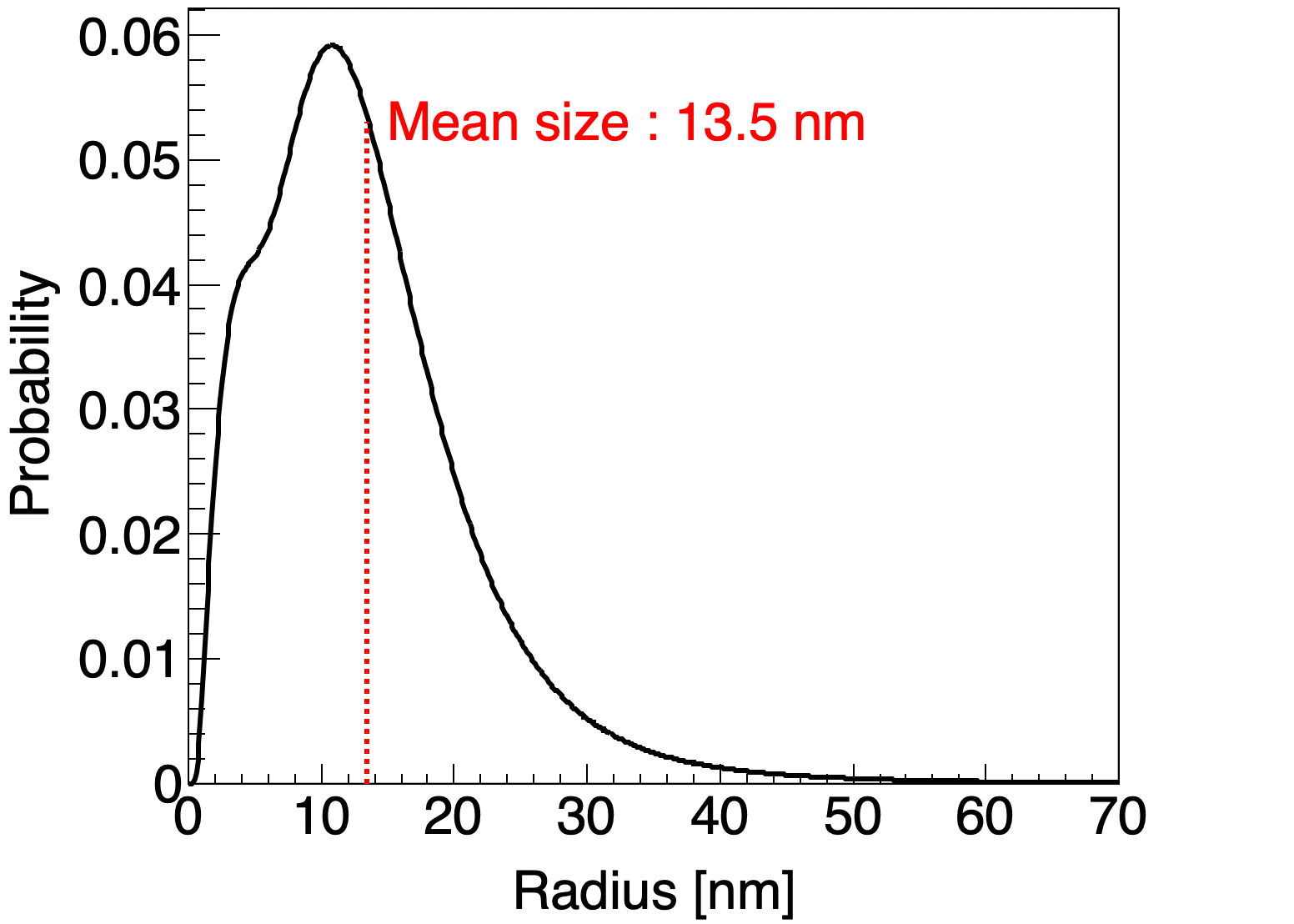}
    \caption{Particle size distribution of $\mathrm{V_{RF2}}$ sample powder obtained by SAXS data.}
    \label{fig:SAXS_par}
\end{figure}
As shown in Fig.~\ref{fig:SAXS_q}(c), the fitting with a Gaussian distribution was found to be reasonable, and the standard deviation was determined as $\mathrm{0.26\,\%}$.
The particle size distribution of the $\mathrm{V_{RF2}}$ sample powder obtained from the fit is shown in Fig.~\ref{fig:SAXS_par}, and the average particle radius was $\mathrm{13.5\,nm}$.

\subsection{Differential cross section}
\begin{figure}
    \centering 
    \includegraphics[width=0.6\textwidth]{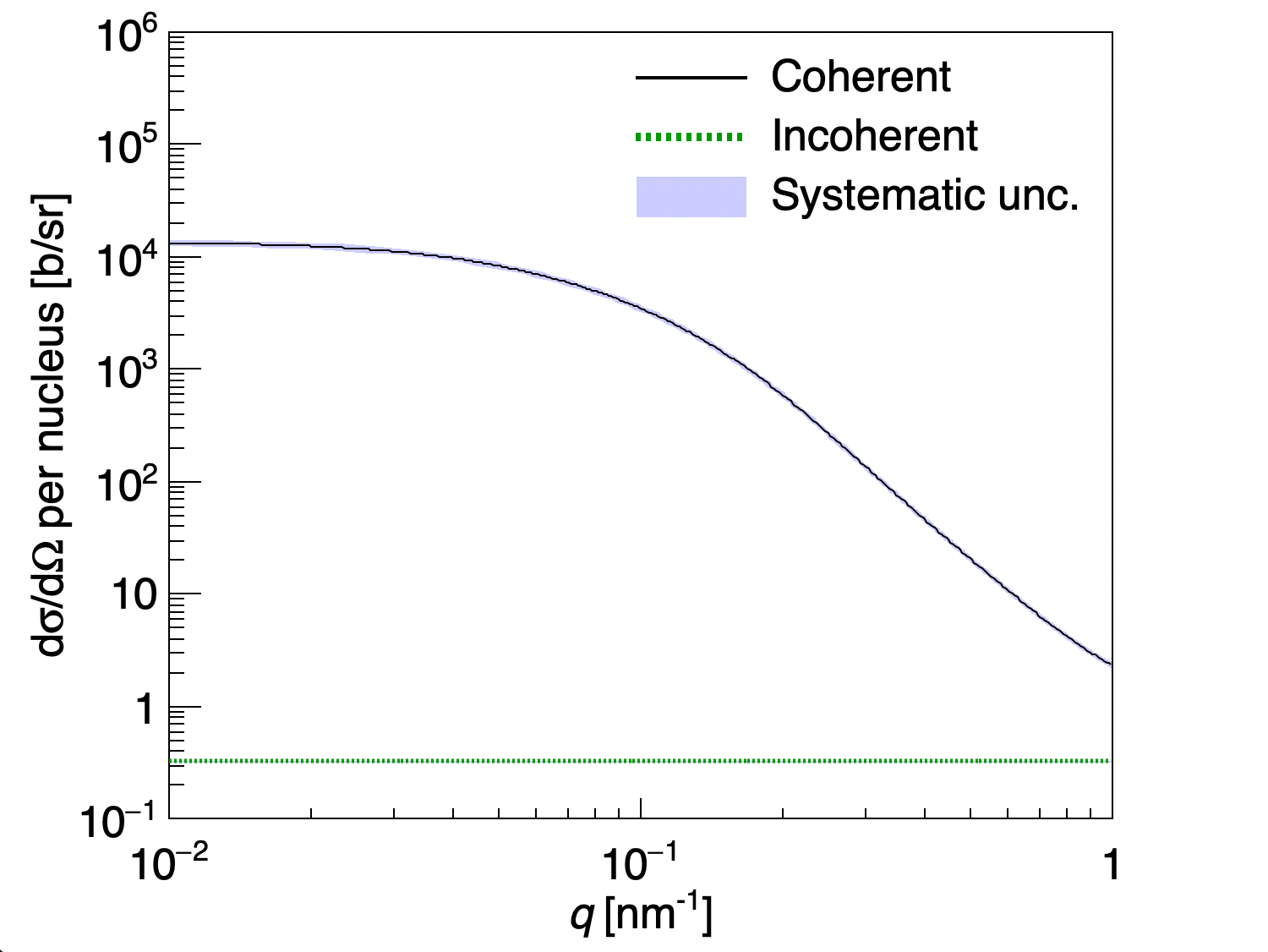}
    \caption{Calculated differential nuclear scattering cross sections for the $\mathrm{V_{RF2}}$ nanoparticle powder as a function of momentum transfer $q$. The coherent and incoherent scattering cross sections were calculated with the parameters listed in Table~\ref{Table:SL_RF2nd}. 
    The particle form factor was calculated assuming spherical particles with a log-normal radius distribution obtained from SAXS measurements. 
     The blue shaded band represents the combined uncertainty arising from variations in elemental composition and uncertainties in the neutron scattering length data.}
    \label{fig:DiffXs_V}
\end{figure}

Figure~\ref{fig:DiffXs_V} shows the differential nuclear scattering cross sections due to coherent and incoherent processes for the $\mathrm{V_{RF2}}$ nanoparticle powder as a function of momentum transfer $q$. 
The calculations were performed using the average coherent scattering length and cross sections derived from the elemental composition in Sec.~\ref{sec:4.1}, together with the experimentally determined particle size distribution from SAXS measurements in Sec.~\ref{sec:4.2}.\\
The coherent nuclear scattering intensity exhibits a broad maximum near $q\approx 0.1\,\mathrm{nm^{-1}}$, corresponding to a characteristic length scale of approximately $\mathrm{10\,nm}$. This region coincides with the optimal momentum transfer range for probing hypothetical Yukawa-type interactions with ranges $\lambda_\mathrm{G}\approx 10\,\mathrm{nm}$, and is readily accessible with existing small-angle neutron scattering instruments. Across the experimentally relevant $q$ range, the coherent nuclear scattering contribution dominates over incoherent scattering by more than four orders of magnitude, indicating that incoherent backgrounds are negligible for real measurements using these nanoparticle targets.\\
The blue shaded band represents the combined uncertainty arising from compositional variations and uncertainties in the neutron scattering length data. Even when these uncertainties are taken into account, the resulting variation in the differential cross section remains smooth and featureless in $q$, and therefore does not introduce spurious structures that could mimic the expected signal from short-range new interactions. Since photon is massless, the $q$ distribution of SAXS is not affected by the hypothetical Yukawa interaction. Therefore, a pure size distribution can be obtained from the q distribution of SAXS. Uncertainties associated with the particle size distribution were evaluated from the reproducibility of the SAXS fits and amount to approximately $\mathrm{0.26\,\%}$, which is negligible compared with the systematic uncertainties of the nuclear scattering length data.\\
These results demonstrate that vanadium-based nanoparticle targets fabricated in this work exhibit well-controlled and predictable nuclear scattering properties over the relevant momentum transfer range, satisfying the key experimental requirements for precision neutron scattering measurements aimed at searching for new gravity-like short-range interactions.

\section{Summary}\label{}
In this work, we developed a nanoparticle sample which is suitable as the target material of experimental search for new short-range gravity-like interactions by means of the small-angle coherent neutron scattering.
Vanadium-based nanoparticles developed in the present work were found to satisfy the requirements from the real experiment as summarized below.
The nanoparticles were fabricated using a RF thermal plasma method, which significantly reduced contamination compared to mechanical milling processes. Comprehensive characterization using SEM, elemental analysis, and SAXS confirmed that the particles had a well-defined particle size distribution and controlled chemical composition. The average coherent nuclear scattering length was calculated to be $\mathrm{0.719(23)\,fm}$, whose absolute value remains comparable to that of natural vanadium.
The calculated differential neutron scattering cross sections show that coherent nuclear scattering dominates over incoherent scattering and exhibits a smooth and predictable $q$ dependence. This behavior meets the essential requirements for neutron scattering and interferometry experiments, where well-characterized and minimized nuclear background radiation is crucial.
In the future, further reduction and control of the average coherent scattering length may be possible by utilizing the hydrogen absorption properties of vanadium. 
Since the coherent nuclear scattering length of hydrogen is a negative value $\mathrm{-3.7390\,fm}$, controlling the amount of hydrogen incorporated into vanadium nanoparticles shows future possibilities for tuning the average nuclear scattering length and achieving a scattering state with zero scattering probability.
In this case, the average coherent nuclear scattering length of the fabricated vanadium nanoparticles is $\mathrm{0.719(23)\,fm}$, while the calculated average coherent nuclear scattering length when incorporating $\mathrm{0.43(1)\,wt\%}$ hydrogen is $\mathrm{0.01(2)\,fm}$. Such tunability will greatly enhance the flexibility of nanoparticle targets in future neutron experiments, including the exploration of new short-range gravity-like interactions. The present work established the production method of customized low-scattering nanoparticle targets, which will be crucial for low-background slow-neutron experiments.

\section*{Acknowledgment}
This research was supported by the Japan Society for the Promotion of Science (JSPS) KAKENHI Grant Numbers 19H01927 and 23K22502. We would like to thank the members of the Neutron Optical and Physics (NOP) group for supporting this work. We would like to thank Prof. William Michael Snow of Indiana University and Prof. Albert Young of the University of North Carolina for their valuable discussions on this work and fundamental neutron research. We would also like to express our sincere gratitude to Prof. Hiroyasu Ejiri for his generous support and for giving us the opportunity to present this research internationally through the support of the Nogami Fund.
We would also like to thank Dr. Masato Tamura from RCNP, Graduate School of Science, Osaka University, Mr. Yoshifumi Sakai from Nissin Engineering Co., Ltd., and Mr. Nobuo Tsukihara from Aisin Nanotechnologies Co., Ltd. for their assistance with the target prototype experiments.
We would also like to express our sincere gratitude to Ms. Nao Eguchi and Mr. Yosuke Murakami from the Institute of Scientific and Industrial Research, Osaka University, for their invaluable assistance with the sample analysis process and measurements.

\bibliographystyle{plain}

 \bibliography{Bib}
 
\end{document}